\documentclass[english,american]{article}
\usepackage[T1]{fontenc}
\usepackage{geometry}
\usepackage{fancyhdr}
\usepackage{color}
\usepackage{babel}
\usepackage{verbatim}
\usepackage{prettyref}
\usepackage{units}
\usepackage{textcomp}
\usepackage{amsmath}
\usepackage{graphicx}
\usepackage{esint}
\usepackage[unicode=true,
 bookmarks=true,bookmarksnumbered=false,bookmarksopen=false,
 breaklinks=true,pdfborder={0 0 0},backref=false,colorlinks=false]
 {hyperref}
\hypersetup{pdftitle={Limits to network scaling},
 pdfauthor={van Albada, Helias, Diesmann},
 pdfsubject={correlated neural networks},
 colorlinks=true,linkcolor=blue,citecolor=blue,urlcolor=blue,filecolor=blue}
\usepackage{breakurl}

\makeatletter

\providecommand{\tabularnewline}{\\}

\newcommand{\lyxaddress}[1]{
\par {\raggedright #1
\vspace{1.4em}
\noindent\par}
}

\newrefformat{alg}{Alg.~\ref{#1}}
\newrefformat{fig}{Fig.~\ref{#1}}
\newrefformat{tab}{Table ~\ref{#1}}
\newrefformat{sec}{\bf "\ref{#1}"}
\newrefformat{sub}{\bf"\ref{#1}"}
\newrefformat{apx}{Appendix~\ref{#1}}

\newrefformat{eq}{Eq.~(\ref{#1})}
\newrefformat{eqs}{Eqs.~(\ref{#1})}

\renewenvironment{abstract}
{\noindent{\normalfont\large\textbf{Abstract}\pdfbookmark[1]{Abstract}{AbstractPage}}\par\vspace{0.5\baselineskip}}
{\par}

\renewcommand{\@seccntformat}[1]{%
\csname the#1\endcsname\hspace{0.5em}}

\renewcommand{\section}{\@startsection
{section}%
{1}%
{0mm}%
{-\baselineskip}%
{0.5\baselineskip}%
{\normalfont\large\bfseries}}

\renewcommand{\subsection}[1]{\ssubsection{#1.}}
\newcommand{\ssubsection}{\@startsection
{subsection}%
{2}%
{1em}%
{-\baselineskip}%
{-\fontdimen2\font plus -\fontdimen3\font minus -\fontdimen4\font}%
{\normalfont\bfseries}}

\renewcommand{\subsubsection}[1]{\sssubsection{#1.}}
\newcommand{\sssubsection}{\@startsection
{subsubsection}%
{3}%
{1em}%
{-\baselineskip}%
{-\fontdimen2\font plus -\fontdimen3\font minus -\fontdimen4\font}%
{\normalfont\itshape}}

\usepackage{footmisc}

\renewcommand{\@makecaption}[2]{%
{\parbox[t]{\linewidth}{%
\normalsize\renewcommand{\baselinestretch}{1.0}\normalsize
\vspace{2mm}
\textbf{#1:} #2
}}}

\usepackage{xspace}

\usepackage{xcolor}
\definecolor{lightgray}{gray}{0.9}

\usepackage{listings}
\makeatletter
\renewcommand\@biblabel[1]{#1.}
\makeatother

\makeatother

\begin{document}
\global\long\def\second{\,\mathrm{s}}
\global\long\def\ms{\,\mathrm{ms}}

\let\oldnameref\nameref \renewcommand{\nameref}[1]{\textbf{``\oldnameref{#1}''}}

\begin{titlepage}\thispagestyle{empty}\setcounter{page}{0}\pdfbookmark[1]{Title}{TitlePage}

\begin{center}
\textbf{\Large{}Scalability of asynchronous networks is limited by
one-to-one mapping between effective connectivity and correlations}
\par\end{center}{\Large \par}

\begin{center}
\textbf{\large{}Sacha Jennifer van Albada$^{1}$, Moritz Helias$^{1}$, and Markus Diesmann$^{1,2,3}$}\vspace{3cm}

\par\end{center}

\lyxaddress{$^{1}$\parbox[t]{15cm}{Institute of Neuroscience and Medicine (INM-6)
and Institute for Advanced Simulation (IAS-6),\\
J{\"u}lich Research Centre and JARA, J{\"u}lich, Germany}\\[3mm]$^{2}$\parbox[t]{15cm}{Department
of Psychiatry, Psychotherapy and Psychosomatics, Medical Faculty,
RWTH Aachen University, Germany}\\[3mm]$^{3}$\parbox[t]{15cm}{Department
of Physics, Faculty 1, RWTH Aachen University, Germany}\\[3mm]\vfill{}
}

\noindent\\
Correspondence to:\hspace{1em}\parbox[t]{11cm}{Dr.\ Sacha van Albada\\

\newlength{\myw}\settowidth{\myw}{fax:\ }\makebox[\myw][l]{tel: +49-2461-61-1944}

\href{mailto:s.van.albada@fz-juelich.de}{s.van.albada@fz-juelich.de}

}\end{titlepage}
\begin{abstract}
Network models are routinely downscaled compared to nature in terms
of numbers of nodes or edges because of a lack of computational resources,
often without explicit mention of the limitations this entails. While
reliable methods have long existed to adjust parameters such that
the first-order statistics of network dynamics are conserved, here
we show that limitations already arise if also second-order statistics
are to be maintained. The temporal structure of pairwise averaged
correlations in the activity of recurrent networks is determined by
the effective population-level connectivity. We first show that in
general the converse is also true and explicitly mention degenerate
cases when this one-to-one relationship does not hold. The one-to-one
correspondence between effective connectivity and the temporal structure
of pairwise averaged correlations implies that network scalings should
preserve the effective connectivity if pairwise averaged correlations
are to be held constant. Changes in effective connectivity can even
push a network from a linearly stable to an unstable, oscillatory
regime and vice versa. On this basis, we derive conditions for the
preservation of both mean population-averaged activities and pairwise
averaged correlations under a change in numbers of neurons or synapses
in the asynchronous regime typical of cortical networks. We find that
mean activities and correlation structure can be maintained by an
appropriate scaling of the synaptic weights, but only over a range
of numbers of synapses that is limited by the variance of external
inputs to the network. Our results therefore show that the reducibility
of asynchronous networks is fundamentally limited. 
\end{abstract}

\section{Introduction}

\lhead{Limits to network scaling}

\rhead{van Albada et al.}

While many aspects of brain dynamics and function remain unexplored,
the numbers of neurons and synapses in a given volume are well known,
and as such constitute basic parameters that should be taken seriously.
Despite rapid advances in neural network simulation technology and
increased availability of computing resources \cite{Albada14}, memory
and time constraints still lead to neuronal networks being routinely
downscaled both on traditional architectures \cite{Helias12_26} and
in systems dedicated to neural network simulation \cite{Khan08_2849}.
As synapses outnumber neurons by a factor of $10^{3}-10^{5}$, these
constitute the main constraint on network size. Computational capacity
ranges from a few tens of millions of synapses on laptop or desktop
computers, or on dedicated hardware when fully exploited \cite{Bruederle11_263,Sharp14},
to $10^{12}-10^{13}$ synapses on supercomputers \cite{Kunkel14_78}.
This upper limit is still about two orders of magnitude below the
full human brain, underlining the need for downscaling in computational
modeling. In fact, any brain model that approximates a fraction of
the recurrent connections as external inputs is in some sense downscaled:
the missing interactions need to be absorbed into the network and
input parameters in order to obtain the appropriate statistics. Unfortunately,
the implications of such scaling are usually not investigated.

\begin{figure*}
\centering{}\includegraphics{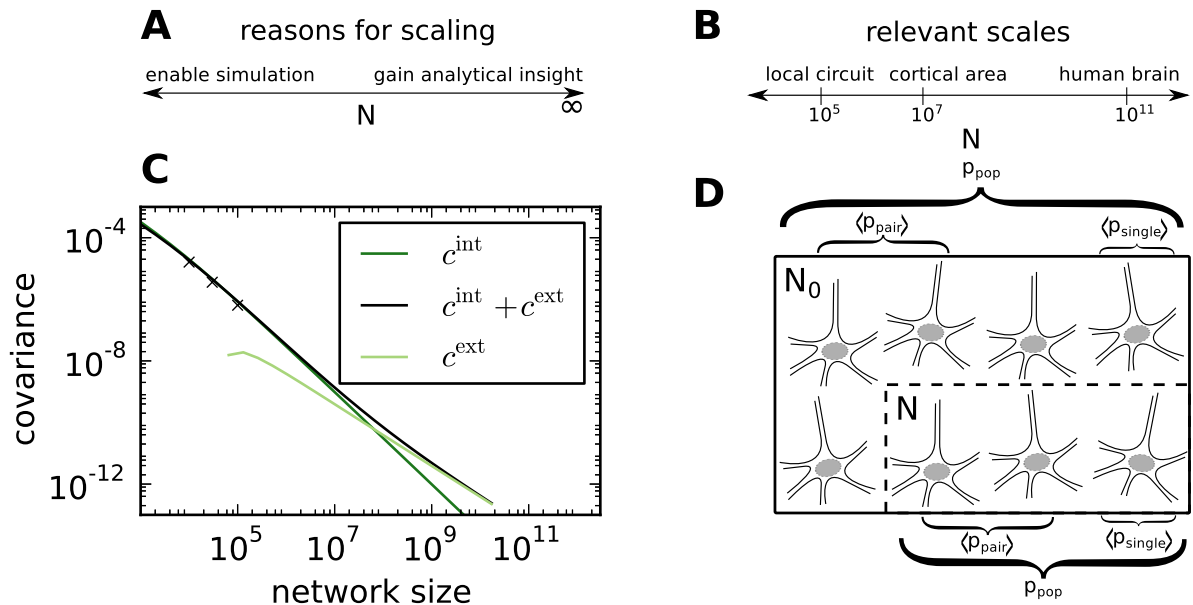}\protect\caption{Framework for neural network scaling. \textbf{A} Downscaling facilitates
simulations, while taking the $N\to\infty$ limit often affords analytical
insight. \textbf{B} Relevant scales. The local cortical microcircuit
containing roughly $10^{5}$ neurons is the smallest network where
the majority of the synapses ($\sim10^{4}$ per neuron) can be represented
using realistic connection probabilities ($\sim0.1$). \textbf{C}
Results for the $N\to\infty$ limit may not apply even for large networks.
In this example, analytically determined intrinsic and extrinsic contributions
to correlations between excitatory neurons are shown. The extrinsic
contribution to the correlation between two neurons arises due common
external input, and the intrinsic contribution due to fluctuations
generated within the network (cf. \cite{Helias14} eq. 24). The intrinsic
contribution falls off more rapidly than the extrinsic contribution,
but nevertheless dominates up to large network sizes, here around
$10^{8}$. The crosses indicate simulation results. Adapted from \cite{Helias14}
Fig. 7. \textbf{D} Scaling transformations may be designed to preserve
average single-neuron or pairwise statistics for selected quantities,
population statistics, or a combination of these. When average single-neuron
and pairwise properties are preserved, the downscaled network of size
$N$ behaves to second order like a subsample of the full network
of size $N_{0}$.}
\label{fig:scaling_framework}
\end{figure*}

The opposite type of scaling, taking the infinite size limit, is sometimes
used in order to simplify equations describing the network (\prettyref{fig:scaling_framework}A).
Although this can lead to valuable insights, real networks in the
human brain often contain on the order of $10^{5}\text{\textminus}10^{7}$
neurons (\prettyref{fig:scaling_framework}B), too few to simplify
certain equations in the limit of infinite size. This is illustrated
in \prettyref{fig:scaling_framework}C using as an example the intrinsic
contribution to correlations due fluctuations generated within the
network, and the extrinsic contribution due to common external inputs
to different neurons in random networks. Although the intrinsic contribution
falls off more rapidly than the extrinsic one, it is the main contribution
up to large network sizes (around $10^{8}$ for the given parameters).
Therefore, taking the infinite size limit and neglecting the intrinsic
contribution leads to the wrong conclusions: The small correlations
in finite random networks cannot be explained by the network activity
tracking the external drive \cite{Renart10_587}, but rather requires
the consideration of negative feedback \cite{Tetzlaff12_e1002596}
that suppresses intrinsically generated and externally imprinted fluctuations
alike \cite{Helias14}. 

Taking the infinite size limit for analytical tractability and downscaling
to make networks accessible by direct simulation are two separate
problems. We concentrate in the remainder of this study on such downscaling,
which is often performed not only in neuroscience \cite{Wilson92_981,Tsodyks95_111,Hill05_1671,Izhikevich08_3593}
but also in other disciplines \cite{Winslow93,Morris97,TenTusscher06,Bisset09}.
Neurons and synapses may either be subsampled or aggregated \cite{Crook12};
here we focus on the former. One intuitive way of scaling is to ensure
that the statistics of particular quantities of interest in the downscaled
network match that of a subsample of the same size from the full network
(\prettyref{fig:scaling_framework}D). Alternatively, it may sometimes
be useful to preserve the statistics of population sums of certain
quantities, for instance population fluctuations. 

We here focus on the preservation of mean population-averaged activities
and pairwise averaged correlations in the activity. We consider both
the size and temporal structure of correlations, but not distributions
of mean activities and correlations across the network. Means and
correlations present themselves as natural quantities to consider,
because they are the first- and second-order and as such the most
basic measures of the dynamics. If it is already difficult to preserve
these measures, it is even less likely that preserving higher-order
statistics will be possible, in view of their higher dimensionality.
However, other choices are possible, for instance maintaining total
input instead of output spike rates \cite{Amit-1997_373}. 

Besides being the most basic dynamical characteristics, means and
correlations of neural activity are biologically relevant. Mean firing
rates are important in many theories of network function \cite{Amit91_275,Gerstner92},
and their relevance is supported by experimental results \cite{Romo99,Ahissar00_302}.
For instance, neurons exhibit orientation tuning of spike rate in
the visual system \cite{Hubel68} and directional tuning in the motor
system \cite{Georgopoulos86_1416}, and sustained rates are implicated
in the working memory function of the prefrontal cortex \cite{Romo99}.
Firing rates have also been shown to be central to pattern learning
and retrieval in highly connected recurrent neural networks \cite{Gerstner92}.
Furthermore, mean firing rates distinguish between states of arousal
and attention \cite{Steriade01_1969,Roelfsema96}, and between healthy
and disease conditions \cite{Albada09_642}. The relevance of correlations
is similarly supported by a large number of findings. They are widely
present; multi-unit recordings have revealed correlated neuronal activity
in various animals and behavioral conditions \cite{Perkel67b,Aertsen89,Kilavik09_12653}.
Pairwise correlations were even shown to capture the bulk of the structure
in the spiking activity of retinal and cultured cortical neurons \cite{Schneidman06_1007}.
They are also related to information processing and behavior. Synchronous
spiking (corresponding to a narrow peak in the cross-correlogram)
has for example been shown to occur in relation to behaviorally relevant
events \cite{Ito11_2482,Riehle97_1950,Vaadia95a}. The relevance of
correlations for information processing is further established by
the fact that they can increase or decrease the signal-to-noise ratio
of population signals \cite{Sompolinsky01a,Zohary94_140}. Moreover,
correlations are important in networks with spike-timing-dependent
plasticity, since they affect the average change in synaptic strengths
\cite{Izhikevich03}. Correspondingly, for larger correlations, stronger
depression is needed for an equilibrium state with asynchronous firing
and a unimodal weight distribution to exist in balanced random networks
\cite{Morrison07_1437}. The level of correlations in neuronal activity
has furthermore been shown to affect the spatial range of local field
potentials (LFPs) effectively sampled by extracellular electrodes
\cite{Linden11_859}. More generally, mesoscopic and macroscopic measures
like the LFP and fMRI depend on interneuronal correlations \cite{Nir07_1275}.
Considering the wide range of dynamical and information processing
properties affected by mean activities and correlations, it is important
that they are accurately modeled.

We allow the number of neurons $N$ and the number of incoming synapses
per neuron $K$ (the in-degree) to be varied independently, generalizing
the common type of scaling where the connection probability is held
constant so that $N$ and $K$ change proportionally. It is well known
that reducing the number of neurons in asynchronous networks increases
correlation sizes in inverse proportion to the network size \cite{Ginzburg94,Amit-1997_373,Vreeswijk98,Hertz10_427,Helias13_023002}.
However, the influence of the number of synapses on the correlations,
including their temporal structure, is less studied. When reducing
the number of synapses, one may attempt to recover aspects of the
network dynamics by adjusting parameters such as the synaptic weights
$J$, the external drive, or neurotransmitter release probabilities
\cite{Tsodyks95_111,Amit-1997_373}. In the present work, spike transmission
is treated as perfectly reliable. We only adjust the synaptic weights
and a combination of the neuronal threshold and the mean and variance
of the external drive to make up for changes in $N$ and $K$. 

\begin{figure*}
\centering{}\includegraphics{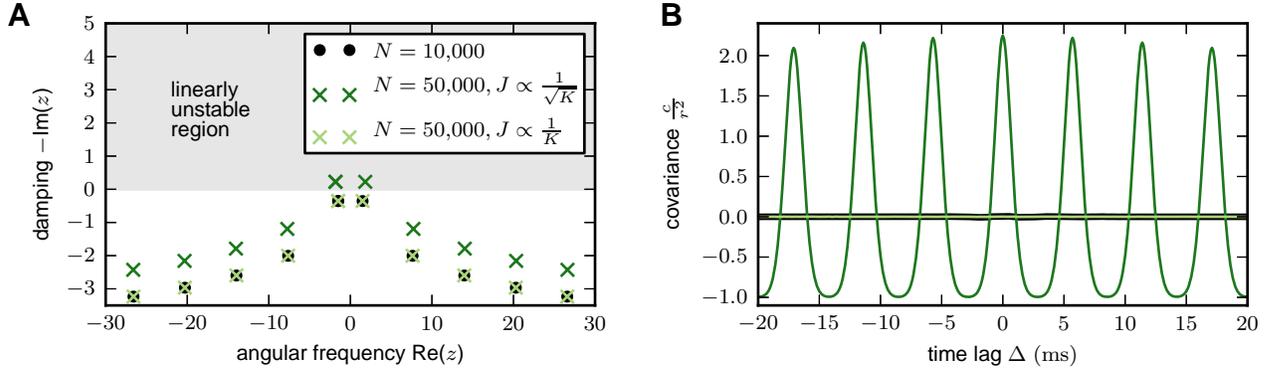}\protect\caption{Transforming synaptic strengths $J$ with the square root of the number
of incoming synapses per neuron $K$ (the in-degree) upon scaling
of network size $N$ changes correlation structure when mean and variance
of the input current are maintained. A reference network of $10,000$
inhibitory leaky integrate-and-fire neurons is scaled up to $50,000$
neurons, fixing the connection probability and adjusting the external
Poisson drive to keep the mean and variance of total (external plus
internal) inputs fixed. Single-neuron parameters and connection
probability are as in \prettyref{tab:spiking_network-1}. Delays are
$1\,\mathrm{ms}$, mean and standard deviation of total inputs are
$15\,\mathrm{mV}$ and $10\,\mathrm{mV}$, respectively, and the reference
network has $J=0.1\,\mathrm{mV}$. Each network is simulated for $50\,\mathrm{s}$.\textbf{
A} Onset of oscillations induced by scaling of network size $N$,
visualized by changes in the poles $z$ of the covariance function
in the frequency domain. $\mathrm{Re}(z)$ determines the frequency
of oscillations and Im($z$) their damping, such that $\mathrm{-Im}(z)>0$
means that small deviations from the fixed-point activity of the network
grow with time {[}cf. \eqref{eq:rate_equation_projected}{]}. The
transformation \textrm{$J\propto\frac{1}{K}$} preserves the poles,
while \textrm{$J\propto\frac{1}{\sqrt{K}}$ }induces a Hopf bifurcation
so that the scaled network is outside the linearly stable regime.
\textbf{B} Covariance in the network where coupling strength $J$
is scaled with the in-degree $K$ matches that in the reference network,
whereas large oscillations appear in the network scaled with $\sqrt{K}$.
Colors as in \textbf{A}.}
\label{fig:hopf}
\end{figure*}

A few suggestions have been made for adjusting synaptic weights to
numbers of synapses. In the balanced random network model, the asynchronous
irregular (AI) firing often observed in cortex is explained by a domination
of inhibition which causes a mean membrane potential below spike threshold,
and sufficiently large fluctuations that trigger spikes \cite{VanVreeswijk98_1321}.
In order to achieve such an AI state for a large range of network
sizes, one choice is to ensure that input fluctuations remain similar
in size, and adjust the threshold or a DC drive to maintain the mean
distance to threshold. As fluctuations are proportional to $J^{2}\, K$
for independent inputs, this suggests the scaling

\begin{equation}
J\propto\frac{1}{\sqrt{K}}\label{eq:Vreeswijk_scaling}
\end{equation}
proposed in \cite{VanVreeswijk98_1321}. Since the mean input to
a neuron is proportional to $J\, K$, \eqref{eq:Vreeswijk_scaling}
leads, all else being equal, to an increase of the population feedback
with $\sqrt{K}$, changing the correlation structure of the network,
as illustrated in \prettyref{fig:hopf} for a simple network of inhibitory
leaky integrate-and-fire neurons (note that in this example we fix
the connection probability). 
This suggests the alternative \cite{Ginzburg94,Hertz10_427,Helias13_023002}
\begin{equation}
J\propto\frac{1}{K},\label{eq:Helias_scaling}
\end{equation}
where now the variance of the external drive needs to be adjusted
to maintain the total input variance onto neurons in the network.

For a given network size $N$ and mean activity level, the size and
temporal structure of pairwise averaged correlations are determined
by the so-called \textit{effective connectivity}, which quantifies
the linear dependence of the activity of each target population on
the activity of each source population. The effective connectivity
is proportional to synaptic strength and the number of synapses a
target neuron establishes with the source population, and additionally
depends on the activity of the target neurons. Effective connectivity
has previously been defined as ``the experiment and time-dependent,
simplest possible circuit diagram that would replicate the observed
timing relationships between the recorded neurons'' \cite{Aertsen90}.
In our analysis we consider the stationary state, but at different
times the network may be in a different state exhibiting a different
effective connectivity. The definition of \cite{Aertsen90} highlights
the fact that identical neural timing relationships can in principle
occur in different physical circuits and vice versa. However, with
a given model of interactions or coupling, the activity may allow
a unique effective connectivity to be derived \cite{Friston11}. We
define effective connectivity in a forward manner with knowledge of
the physical connectivity as well as the form of interactions. We
show in this study that with this model of interactions, and with
independent external inputs, the activity indeed determines a unique
effective connectivity, so that the forward and reverse definitions
coincide. This complements the groundbreaking general insight of \cite{Aertsen90}.

We consider networks of binary model neurons and networks of leaky
integrate-and-fire (LIF) neurons with current-based synapses to investigate
how and to what extent changes in network parameters can be used to
preserve mean population-averaged activities and pairwise averaged
correlations under reductions in the numbers of neurons and synapses.
The parameters allowed to vary are the synaptic weights, neuronal
thresholds, and the mean and variance of the external drive. We apply
and extend the theory of correlations in randomly connected binary
and LIF networks in the asynchronous regime developed in \cite{Ginzburg94,Lindner05_061919,Renart10_587,Pernice11_e1002059,Tetzlaff12_e1002596,Trousdale12_e1002408,Grytskyy13_258,Grytskyy13_131,Helias13_023002,Helias14},
which explains the smallness and structure of correlations experimentally
observed during spontaneous activity in cortex \cite{Okun2008_535,Graupner2013_15075},
and we compare analytical predictions of correlations with results
from simulations. The results are organized as follows. In \nameref{sub:Correlations-uniquely-determine}
we provide an intuitive example that illustrates why the effective
connectivity uniquely determines correlation structure. In \nameref{sub:Correlations-uniquely-determine-general}
we show that this one-to-one relationship generalizes to networks
of several populations apart from degenerate cases. In 
%\textbf{\textcolor{blue}{``Correlation-preserving scaling''}}
\nameref{sub:Correlation-preserving-scaling}
we conclude that, in general, only scalings that preserve the effective
connectivity, such as $J\propto1/K$, are able to preserve correlations.
In \nameref{sub:Limit-to-in-degree} we identify the limits of the
resulting scaling procedure, demonstrating the restricted scalability
of asynchronous networks. 
%\textbf{\textcolor{blue}{``Robustness of correlation-preserving scaling''}}
\nameref{sub:Generalizability}
shows that the scaling $J\propto1/K$ can preserve correlations, within
the identified restrictive bounds, for different networks either adhering
to or deviating from the assumptions of the analytical theory. 
%\textbf{\textcolor{blue}{``Zero-lag correlations in binary network''}}
\nameref{sub:Zero-lag-corr}
investigates how to maintain the instantaneous correlations in a binary
network, while 
%\textbf{\textcolor{blue}{``Symmetric two-population spiking network''}}
\nameref{sub:Symmetric-two-population-spiking}
considers the degenerate case of a connectivity with special symmetries,
in which correlations may be maintained under network scaling without
preserving the effective connectivity. Preliminary results have been
published in abstract form \cite{Albada13_CNS}.

\section{Results}

\subsection{Correlations uniquely determine effective connectivity: a simple
example\label{sub:Correlations-uniquely-determine}}

In this section we give an intuitive one-dimensional example to show
that effective connectivity determines the shapes of the average pairwise
cross-covariances and vice versa. For the following, we first introduce
a few basic quantities. Consider a binary or spiking network consisting
of several excitatory and inhibitory populations with potentially
source- and target-type dependent connectivity. For the spiking networks,
we assume leaky integrate-and-fire (LIF) dynamics with exponential
synaptic currents. The dynamics of the binary and LIF networks are
respectively introduced in \nameref{sub:Binary-network-dynamics}
and \nameref{sub:Spiking-network-dynamics}. We assume irregular network
activity, approximated as Poissonian for the spiking network, with
population means $\nu_{\alpha}$. For the binary network, $\nu=\langle n\rangle$
is the expectation value of the binary variable. For the spiking network,
we absorb the membrane time constant into $\nu$, defining $\nu=\tau_{\mathrm{m}}r$
where $r$ is the firing rate of the population. The external drive
can consist of both a DC component $\mu_{\alpha,\mathrm{ext}}$ and
fluctuations with variance \foreignlanguage{english}{\textrm{$\sigma_{\alpha,\mathrm{ext}}^{2}$}},
provided either by Poisson spikes or a Gaussian current. The working
points of each population, characterized by mean $\mu_{\alpha}$ and
variance $\sigma_{\alpha}^{2}$ of the combined input from within
and outside the network, are given by 
\begin{eqnarray}
\mu_{\alpha} & = & \sum_{\beta}J_{\alpha\beta}K_{\alpha\beta}\nu_{\beta}+\mu_{\alpha,\mathrm{ext}}\label{eq:mu}\\
\sigma_{\alpha}^{2} & = & \sum_{\beta}J_{\alpha\beta}^{2}K_{\alpha\beta}\phi_{\beta}+\sigma_{\alpha,\mathrm{ext.}}^{2}\label{eq:sigma}\\
\text{with}\nonumber \\
\phi & \equiv & \begin{cases}
\left(1-\langle n\rangle\right)\langle n\rangle & \mathrm{for}\:\mathrm{binary}\\
\nu & \mathrm{for}\:\mathrm{LIF}
\end{cases},
\end{eqnarray}
where $J_{\alpha\beta}$ is the synaptic strength from population
$\beta$ to population $\alpha$, and $K_{\alpha\beta}$ is the number
of synapses per target neuron (the in-degree) for the corresponding
projection (we use $\equiv$ in the sense of ``is defined as'').
We call $\sigma_{\alpha,\mathrm{ext}}^{2}$ ``external variance''
in the following, and the remainder ``internal variance''. The
mean population activities are determined by $\mu_{\alpha}$ and $\sigma_{\alpha}$
according to \eqref{eq:binary_mean_activity} and \eqref{eq:siegert}.
Expressions for correlations in binary and LIF networks are given
respectively in \nameref{sub:First-and-second-moments-binary} and
\nameref{sub:LIF_moments}. 

As a one-dimensional example, consider a binary network with a single
population and vanishing transmission delays. The effective connectivity
$\mathbf{W}$ is just a scalar, and the population-averaged autocovariance
$\mathbf{a}$ and cross-covariance $\mathbf{c}$ are functions of
the time lag $\Delta$. We define the population-averaged effective
connectivity as
\begin{equation}
W=w(J,\mu,\sigma)K,
\end{equation}
where $w(J,\mu,\sigma)$ is an effective synaptic weight that depends
on the mean $\mu$ \eqref{eq:mu} and the variance $\sigma^{2}$ \eqref{eq:sigma}
of the input. For LIF networks, $w=\partial r_{\mathrm{target}}/\partial r_{\mathrm{source}}$
is defined via \eqref{eq:LIF_kernel} and can be obtained as the derivative
of \eqref{eq:siegert}. Note that we treat the effective influence
of individual inputs as independent. A more accurate definition of
the population-level effective connectivity, beyond the scope of this
paper, could be obtained by also considering combinations of inputs
in the sense of a Volterra series \cite{Bronstein99}. When the dependence
of $w$ on $J$ is linearized, the effective connectivity can be written
as 
\begin{equation}
W=S(\mu,\sigma)JK,
\end{equation}
where the susceptibility $S(\mu,\sigma)$ measures to linear order
the effect of a unit input to a neuron on its outgoing activity. In
our one-dimensional example, $W$ quantifies the self-influence of
an activity fluctuation back onto the population. Expressed in these
measures, the differential equation \eqref{eq:corr_differential}
for the covariance function takes the form
\begin{equation}
\frac{\tau}{1-W}\frac{d}{d\Delta}c(\Delta)=-c(\Delta)+\frac{W}{1-W}\frac{a(\Delta)}{N},\label{eq:simple_corr}
\end{equation}
with initial condition {[}from \eqref{eq:corr_self_consistent_pop}{]}
\begin{equation}
\left(1-W\right)c(0)=\frac{Wa}{N},
\end{equation}
which is solved by
\begin{equation}
c(\Delta)=\frac{a}{N(1-W)}e^{\frac{W-1}{\tau}\Delta}-\frac{a}{N}e^{-\frac{\Delta}{\tau}}.\label{eq:simple_corr_solution}
\end{equation}
 Equation \eqref{eq:simple_corr_solution} shows that the effective
connectivity $W$ together with the time constant $\tau$ of the neuron
(which we assume fixed under scaling) determines the temporal structure
of the correlations. Furthermore, since a sum of exponentials cannot
equal a sum of exponentials with a different set of exponents, the
temporal structure of the correlations uniquely determines $W$. Hence
we see that there is a one-to-one correspondence between $W$ and
the correlation structure if the time constant $\tau$ is fixed, which
implies that preserving correlation structure under a reduction in
the in-degrees $K$ requires adjusting the effective synaptic weights
$w(J,\mu,\sigma)$ such that the effective connectivity $W$ is maintained.
If, in addition, the mean activity $\langle n\rangle$ is kept constant
this also fixes the variance $a=\langle n\rangle(1-\langle n\rangle)$.
Equation \eqref{eq:simple_corr_solution} shows that, under these
circumstances with $W$ and $a$ fixed, correlation sizes are determined
by $N$.

\subsection{Correlations uniquely determine effective connectivity: the general
case\label{sub:Correlations-uniquely-determine-general}}

More generally, networks consist of several neural populations each
with different dynamic properties and with population-dependent transmission
delays. Since this setting does not introduce additional symmetries,
intuitively the one-to-one relationship between the effective connectivity
and the correlations should still hold. We here show that, under certain
conditions, this is indeed the case. 

Instead of considering the covariance matrix in the time domain, for
population-dependent dynamic properties we find it convenient to stay
in the frequency domain. The influence of a fluctuating input on
the output of the neuron can to lowest order be described by the transfer
function $H(\omega)$. This quantity measures the amplitude and phase
of the modulation of the neuronal activity given that the neuron receives
a small sinusoidal perturbation of frequency $\omega$ in its input.
The transfer function depends on the mean $\mu$ \eqref{eq:mu} and
the variance $\sigma^{2}$ \eqref{eq:sigma} of the input to the neuron.
We here first consider LIF networks; in the Supporting Information
we show how the results carry over to the binary model. 

In \nameref{sub:LIF_moments}, we give the covariance matrix including
the autocovariances in the frequency domain,\textrm{ }\foreignlanguage{english}{\textrm{$\bar{\mathbf{C}}(\omega)=\mathbf{C}(\omega)+\mathbf{A}(\omega)$,}}
as
\begin{equation}
\mathbf{\bar{\mathbf{C}}}(\omega)=\left(\mathbf{1}\!\!\mathbf{1}-\mathbf{M}(\omega)\right)^{-1}\mathbf{A}\left(\mathbf{1}\!\!\mathbf{1}-\mathbf{M}^{T}(-\omega)\right)^{-1},\label{eq:LIF_corr_freq}
\end{equation}
where $\mathbf{M}$ has elements $H_{\alpha\beta}(\omega)W_{\alpha\beta}$.
If $\mathbf{\bar{\mathbf{C}}}(\omega)$ is invertible, we can expand
the inverse of \eqref{eq:LIF_corr_freq} to obtain 

\textrm{
\begin{eqnarray}
\bar{C}_{\alpha\beta}^{-1}(\omega) & = & \sum_{\gamma}\left(\mathbf{\mathbf{1}\!\!\mathbf{1}_{\alpha\gamma}}-M_{\gamma\alpha}(-\omega)\right)A_{\gamma}^{-1}\left(\mathbf{\mathbf{1}\!\!\mathbf{1}}_{\gamma\beta}-M_{\gamma\beta}(\omega)\right)\nonumber \\
 & = & \delta_{\alpha\beta}\,\left(1-W_{\alpha\alpha}\frac{e^{i\omega d_{\alpha\alpha}}}{1-i\omega\tau_{\alpha}}\right)A_{\alpha}^{-1}\left(1-W_{\alpha\alpha}\frac{e^{-i\omega d_{\alpha\alpha}}}{1+i\omega\tau_{\alpha}}\right)\nonumber \\
 & + & (\delta_{\alpha\beta}-1)\Bigg[\left(1-W_{\alpha\alpha}\frac{e^{i\omega d_{\alpha\alpha}}}{1-i\omega\tau_{\alpha}}\right)A_{\alpha}^{-1}W_{\alpha\beta}\frac{e^{-i\omega d_{\alpha\beta}}}{1+i\omega\tau_{\alpha}}\nonumber \\
 & + & W_{\beta\alpha}\frac{e^{i\omega d_{\beta\alpha}}}{1-i\omega\tau_{\beta}}A_{\beta}^{-1}\left(1-W_{\beta\beta}\frac{e^{-i\omega d_{\beta\beta}}}{1+i\omega\tau_{\beta}}\right)\Bigg]\nonumber \\
 & + & \sum_{\gamma\neq\alpha,\beta}W_{\gamma\alpha}\frac{e^{i\omega d_{\gamma\alpha}}}{1-i\omega\tau_{\gamma}}A_{\gamma}^{-1}W_{\gamma\beta}\frac{e^{-i\omega d_{\gamma\beta}}}{1+i\omega\tau_{\gamma}},
\end{eqnarray}
}where we assumed the transfer function to have the form $H(\omega)=\frac{e^{-i\omega d_{\alpha\beta}}}{1+i\omega\tau_{\alpha}},$
which is often a good approximation for the LIF model \cite{Helias13_023002}.
In the second step we distinguish terms that only contribute on the
diagonal ($\alpha=\beta$), those that only contribute off the diagonal
($\alpha\neq\beta$), and those that contribute in either case. For
$\alpha=\beta$, only the first and last terms contribute, and we
get
\begin{eqnarray}
\bar{C}_{\alpha\alpha}^{-1} & = & A_{\alpha}^{-1}\nonumber \\
 & - & \frac{W_{\alpha\alpha}}{A_{\alpha}}\left(\frac{e^{-i\omega d_{\alpha\alpha}}}{1+i\omega\tau_{\alpha}}+\frac{e^{i\omega d_{\alpha\alpha}}}{1-i\omega\tau_{\alpha}}\right)\nonumber \\
 & + & \sum_{\gamma}\frac{W_{\gamma\alpha}^{2}A_{\gamma}^{-1}}{1+\omega^{2}\tau_{\gamma}^{2}}.\label{eq:C_inv_LIF_alpha_alpha}
\end{eqnarray}
If we want to preserve $\bar{\mathbf{C}}$, this fixes $A_{\alpha}$
and thereby also $W_{\alpha\alpha}$, since it multiplies terms with
unique $\omega$-dependence. For $\alpha\neq\beta$, we obtain
\begin{eqnarray}
\bar{C}_{\alpha\beta}^{-1} & = & \frac{W_{\alpha\beta}}{A_{\alpha}}e^{-i\omega d_{\alpha\beta}}\left(-\frac{1}{1+i\omega\tau_{\alpha}}+W_{\alpha\alpha}\frac{e^{i\omega d_{\alpha\alpha}}}{1+\omega^{2}\tau_{\alpha}^{2}}\right)\nonumber \\
 & + & \frac{W_{\beta\alpha}}{A_{\beta}}e^{i\omega d_{\beta\alpha}}\left(-\frac{1}{1-i\omega\tau_{\beta}}+W_{\beta\beta}\frac{e^{-i\omega d_{\beta\beta}}}{1+\omega^{2}\tau_{\beta}^{2}}\right)\nonumber \\
 & + & \sum_{\gamma\neq\alpha,\beta}\frac{W_{\gamma\alpha}W_{\gamma\beta}}{A_{\gamma}}\frac{e^{i\omega\left(d_{\gamma\alpha}-d_{\gamma\beta}\right)}}{1+\omega^{2}\tau_{\gamma}^{2}}.\label{eq:C_inv_LIF_alpha_beta}
\end{eqnarray}
With $A_{\alpha}$ fixed, this additionally fixes $W_{\alpha\beta}$,
in view of the unique $\omega$-dependence it multiplies. 

Since $\mathbf{C}(\omega)=\bar{\mathbf{C}}(\omega)-\mathbf{A}$, a
constraint on $\mathbf{A}$ necessary for preserving $\bar{\mathbf{C}}(\omega)$
may not translate into the same constraint when we only require the
cross-covariances $\mathbf{C}(\omega)$ to be preserved. However,
$\mathbf{C}(\omega)$ and $\bar{\mathbf{C}}(\omega)$ have identical
$\omega$-dependence, as they differ only by constants on the diagonal
(approximating autocorrelations as delta functions in the time domain
\cite{Helias13_023002}). To derive conditions for preserving \foreignlanguage{english}{\textrm{$\mathbf{C}(\omega)$}},
we therefore ignore the constraint on $\mathbf{A}$ but still require
the $\omega$-dependence to be unchanged. A potential transformation
leaving the $\omega$-dependent terms in both \eqref{eq:C_inv_LIF_alpha_alpha}
and \eqref{eq:C_inv_LIF_alpha_beta} unchanged is $A_{\alpha}\to kA_{\alpha}$,
$W_{\alpha\beta}\to kW_{\alpha\beta}$, $W_{\alpha\alpha}\to kW_{\alpha\alpha}$,
but this only works if $\tau_{\alpha}=\tau_{\gamma}$, $d_{\alpha\alpha}-d_{\alpha\beta}=d_{\gamma\alpha}-d_{\gamma\beta}$
for some $\gamma$, and if the terms for the corresponding $\gamma$
are also transformed to offset the change in $W_{\alpha\alpha}W_{\alpha\beta}A_{\alpha}^{-1}$;
or if some of the entries of $\mathbf{W}$ vanish. The $\omega$-dependence
of $\bar{\mathbf{C}}$ and \foreignlanguage{english}{\textrm{$\mathbf{C}$}}
would otherwise change, showing that, at least in the absence of such
symmetries in the delays or time constants, or zeros in the effective
connectivity matrix (i.e., absent connections at the population level,
or inactive populations), there is a one-to-one relationship between
covariances and effective connectivity. Hence, preserving the covariances
requires preserving $\mathbf{A}$ and $\mathbf{W}$ except in degenerate
cases. Note that the autocovariances and hence the firing rates can
be changed while affecting only the size but not the shape of the
correlations, but that the correlation shapes determine $\mathbf{W}$.

Even in case of identical transfer functions across populations, including
in particular equal transmission delays and identical $\tau$, the
one-to-one correspondence between effective connectivity and correlations
can be demonstrated except for a narrower set of degenerate cases.
The argument for $d=0$ proceeds in the time domain along the same
lines as \nameref{sub:Correlations-uniquely-determine}, using the
fact that for a population-independent transfer function, the correlations
can be expressed in terms of the eigenvalues and eigenvectors of the
effective connectivity matrix (cf. \nameref{sub:First-and-second-moments-binary}
and \nameref{sub:LIF_moments}). For general delays, a derivation
in the frequency domain can be used. Through these arguments, we show
in the Supporting Information that the one-to-one correspondence holds
at least if $\mathbf{W}$ is diagonalizable and has no eigenvalues
that are zero or degenerate. 

\begin{table}
\centering{}%
\begin{tabular}{|c|c|c|}
\hline 
numbers of neurons & $N_{\mathrm{e}},\, N_{\mathrm{i}}$ & $5000,\,5000$\tabularnewline
\hline 
neuron time constant & $\tau$ & $10\,\mathrm{ms}$ \tabularnewline
\hline 
threshold & $\theta$ & $0$\tabularnewline
\hline 
connection probabilities & $p_{\mathrm{ee}},\, p_{\mathrm{ei}},\, p_{\mathrm{ie}},\, p_{\mathrm{ii}}$ & $0.1,\,0.2,\,0.3,\,0.4$\tabularnewline
\hline 
transmission delay & $d$ & $0.1\,\mathrm{ms}$ \tabularnewline
\hline 
synaptic weights & $J_{\mathrm{ee}},\, J_{\mathrm{ei}},\, J_{\mathrm{ie}},\, J_{\mathrm{ii}}$ & $3,\,-5,\,3,\,-6$\tabularnewline
\hline 
mean external drive & $m_{\mathrm{ex}},m_{\mathrm{ix}}$ & $50,\,40$\tabularnewline
\hline 
SD of external drive & $\sigma_{\mathrm{ex}},\sigma_{\mathrm{ix}}$ & $60,\,50$\tabularnewline
\hline 
\end{tabular}\protect\caption{Parameters of the asymmetric binary network.}
\label{tab:binary_network-1}
\end{table}

\subsection{Correlation-preserving scaling\label{sub:Correlation-preserving-scaling}}

\begin{figure*}
\centering{}\includegraphics{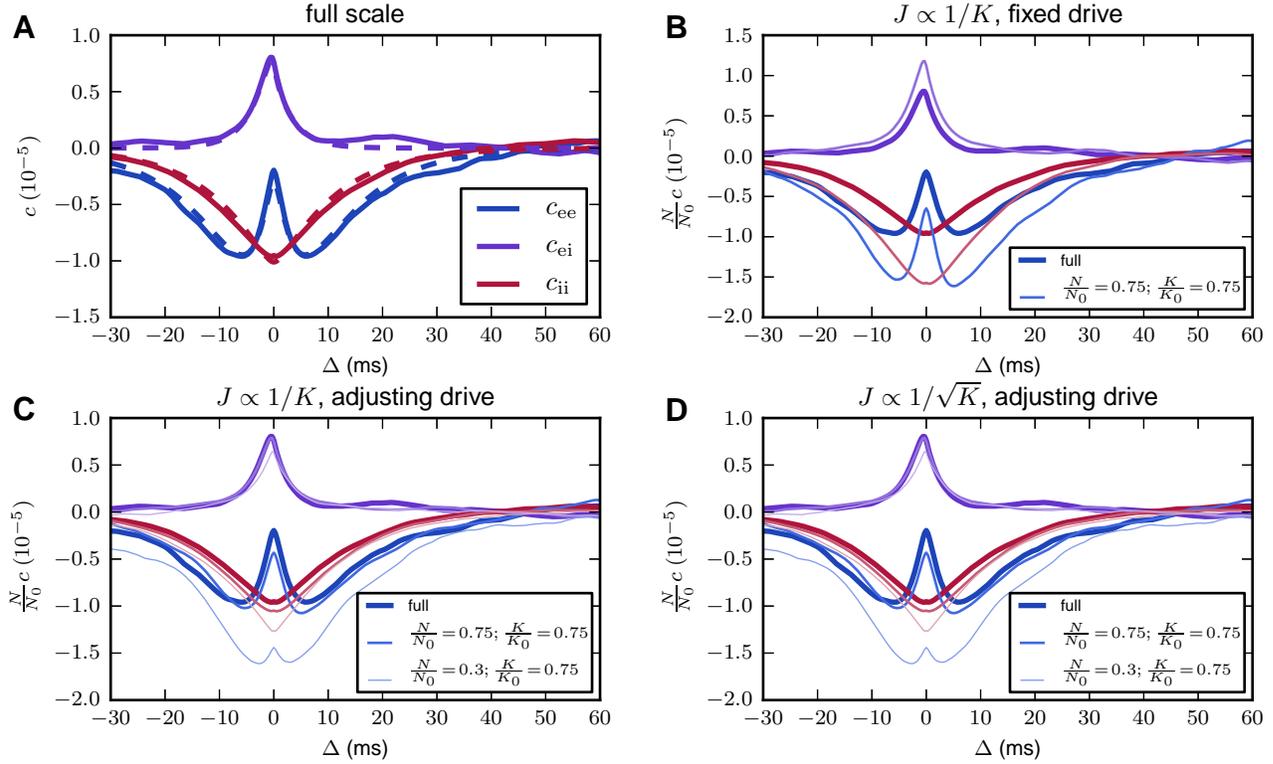}\protect\caption{Correlations from theory and simulations for a two-population binary
network with asymmetric connectivity. \textbf{A} Average pairwise
cross-covariances from simulations (solid curves) and \eqref{eq:binary_cov_matrix_vs_time}
(dashed curves). \textbf{B} Naive scaling with $J\propto1/K$ but
without adjustment of the external drive changes the correlation structure.
\textbf{C} With an appropriate adjustment of the external drive (\foreignlanguage{english}{\textrm{$\sigma_{\mathrm{ex}}=53.4,\;\sigma_{\mathrm{ix}}=17.7$}}),
scaling synaptic weights as $J\propto1/K$ is able to preserve correlation
structure as long as $N$ and $K$ are reduced by comparable factors.
\textbf{D} The same holds for $J\propto1/\sqrt{K}$ (\foreignlanguage{english}{\textrm{$\mu_{\mathrm{ex}}=43.3,\;\mu_{\mathrm{ix}}=34.6,\;\sigma_{\mathrm{ex}}=46.2,\;\sigma_{\mathrm{ix}}=15.3$}}),
but the susceptibility $S$ is increased by about $20\%$ already
for $N=0.75\, N_{0}$ in this case. In \textbf{B}, \textbf{C}, and
\textbf{D}, results of simulations are shown. The curves in\textbf{
C }and\textbf{ D} are identical because internal inputs, the standard
deviation of the external drive, and the distance to threshold due
to the DC component of the drive in\textbf{ D} are exactly $\sqrt{\frac{K}{K_{0}}}$
times those in \textbf{C}. Hence, identical realizations of the random
numbers for the connectivity and the Gaussian external drive cause
the total inputs to the neurons to exceed the threshold at exactly
the same points in time in the two simulations. The simulated time
is $30\protect\second$, and the population activity is sampled at
a resolution of $0.3\protect\ms$.\label{fig:binary_scaling_asymm}}
\end{figure*}

If the working point $(\mu,\sigma)$ is maintained, the one-to-one
correspondence between the effective connectivity and the correlations
implies that requiring unchanged average covariances leaves no freedom
for network scaling except for a possible trade-off between in-degrees
and synaptic weights. In the linear approximation \foreignlanguage{english}{$w(J,\mu,\sigma)=S(\mu,\sigma)JK$},
this trade-off is $J\propto1/K$.

When this scaling is implemented naively without adjusting the external
drive to recover the original working point, the covariances change,
as illustrated in \prettyref{fig:binary_scaling_asymm}B for a two-population
binary network with parameters given in \prettyref{tab:binary_network-1}.
The results of $J\propto1/K$ scaling with appropriate adjustment
of the external drive are shown in \prettyref{fig:binary_scaling_asymm}C.
 The scaling shown in \prettyref{fig:binary_scaling_asymm}B also
increases the mean activities (E: from $0.16$ to $0.23$, I: from
$0.07$ to $0.11$), whereas that in \prettyref{fig:binary_scaling_asymm}C
preserves them.

If one relaxes the constraint on the working point while still requiring
mean activities to be preserved, the network does have additional
symmetries due to the fact that only some combination of $\mu$ and
$\sigma$ needs to be fixed, rather than each of these separately.
This combination is more easily determined for binary than for LIF
networks, for which the mean firing rates depend on $\mu$ and $\sigma$
in a complex manner {[}cf. \eqref{eq:siegert}{]}. When the derivative
of the gain function is narrow (e.g., having zero width in the case
of the Heaviside function used here) compared to the input distribution,
the mean activities of binary networks depend only on $(\mu-\theta)/\sigma$
\cite{Helias14}. Changing $\sigma$ while preserving $(\mu-\theta)/\sigma$
leads for a Heaviside gain function to a new susceptibility $S^{\prime}=(\sigma/\sigma^{\prime})S$
{[}cf. \eqref{eq:susceptibility}{]}. For constant $K$, if the standard
deviation of the external drive is changed proportionally to the internal
standard deviation, we have $\sigma\propto J$ and thus $J^{\prime}S^{\prime}=JS$,
implying an insensitivity of the covariances to the synaptic weights
$J$ \cite{Grytskyy13_258}. In particular, this symmetry applies
in the absence of an external drive. When $K$ is altered, this choice
for adjusting the external drive causes the covariances to change.
However, adjusting the external drive such that $\sigma^{\prime}/\sigma=(J^{\prime}K^{\prime})/(JK)$,
the change in $S$ is countered to preserve $\mathbf{W}$ and correlations.
This is illustrated in \prettyref{fig:binary_scaling_asymm}D for
$J\propto1/\sqrt{K}$, which is another natural choice, as it preserves
the internal variance if one ignores the typically small contribution
of the correlations to the input variance (\cite{Helias14} Fig. 3D
illustrates the smallness of this contribution for an example network).
This is only one of a continuum of possible scalings preserving mean
activities and covariances (within the bounds described in the following
section) when the working point and hence the susceptibility are allowed
to change.

\subsection{Limit to in-degree scaling\label{sub:Limit-to-in-degree}}

We now show that both the scaling $J\propto1/K$ for LIF networks
(for which we do not consider changes to the working point, as analytic
expressions for countering these changes are intractable), and correlation-preserving
scalings for binary networks (where we allow changes to the working
point that preserve mean activities) are applicable only up to a limit
that depends on the external variance. 

For the binary network, assume a generic scaling $K^{\prime}=\kappa K$,
$J^{\prime}=\iota J$ and a Heaviside gain function. We denote variances
due to inputs from within the network and due to the external drive
respectively by $\sigma_{\mathrm{int}}^{2}$ and $\sigma_{\mathrm{ext}}^{2}.$
The preservation of the mean activities implies $S^{\prime}=(\sigma/\sigma^{\prime})S$
as above, where $\sigma^{2}=\sigma_{\mathrm{\mathrm{ext}}}^{2}+\sigma_{\mathrm{int}}^{2}$.
To keep $SJK$ fixed we thus require

\begin{eqnarray}
\sigma_{\mathrm{int}}^{2}\,^{\prime}+\sigma_{\mathrm{ext}}^{2}\,^{\prime} & = & \left(\iota\kappa\right)^{2}\left(\sigma_{\mathrm{int}}^{2}+\sigma_{\mathrm{ext}}^{2}\right)\nonumber \\
\sigma_{\mathrm{ext}}^{2}\,^{\prime} & = & \iota^{2}\kappa\left[\left(\kappa-1\right)\sigma_{\mathrm{int}}^{2}+\kappa\sigma_{\mathrm{ext}}^{2}\right],\label{eq:generic_scaling}
\end{eqnarray}
where we have used $\sigma_{\mathrm{int}}^{\prime}\approx\iota\sqrt{\kappa}\sigma_{\mathrm{int}}$
in the second line. For $\sigma_{\mathrm{ext}}=0$ this scaling only
works for $\kappa>1$, i.e., increasing instead of decreasing the
in-degrees. More generally, the limit to downscaling occurs when $\sigma_{\mathrm{ext}}^{\prime}=0,$
or
\begin{equation}
\kappa=\frac{\sigma_{\mathrm{int}}^{2}}{\sigma_{\mathrm{int}}^{2}+\sigma_{\mathrm{ext}}^{2}},\label{eq:kappa_min}
\end{equation}
independent of the scaling of the synaptic weights. Thus, larger external
and smaller internal variance before scaling allow a greater reduction
in the number of synapses. The in-degrees of the example network of
\prettyref{fig:binary_scaling_asymm} could be maximally reduced to
$73\%$. Note that $\iota$ could in principle be chosen in a $\kappa$-dependent
manner such that $\sigma_{\mathrm{ext}}^{2}$ is fixed or increased
instead of decreased upon downscaling, namely $\iota\geq\sqrt{\frac{\sigma_{\mathrm{ext}}^{2}}{\kappa^{2}\sigma_{\mathrm{ext}}^{2}+\kappa(\kappa-1)\sigma_{\mathrm{int}}^{2}}}$.
However, \eqref{eq:kappa_min} is still the limit beyond which this
fails, as $\iota$ then diverges at that point.

Note that the limit to the in-degree scaling also implies a limit
on the reduction in the number of neurons for which the scaling equations
derived here allow the correlation structure to be preserved, as a
greater reduction of $N$ compared to $K$ increases the number of
common inputs neurons receive and thereby the deviation from the assumptions
of the diffusion approximation. This is shown by the thin curves in
\prettyref{fig:binary_scaling_asymm}C,D. 

Now consider correlation-preserving scaling of LIF networks. Reduced
$K$ with constant $JK$ does not affect mean inputs {[}cf. \eqref{eq:mu}{]}
but increases the internal variance according to \eqref{eq:sigma}.
To maintain the working point $\left(\mu,\sigma\right)$, it is therefore
necessary to reduce the variance of the external drive. When the drive
consists of excitatory Poisson input, one way of keeping the mean
external drive constant while changing the variance is to add an inhibitory
Poisson drive. With $K^{\prime}=K/\iota$ and $J^{\prime}=\iota J$,
the change in internal variance is $(\iota-1)\sigma_{\mathrm{int}}^{2}$,
where $\sigma_{\mathrm{int}}^{2}$ is the internal variance due to
input currents in the full-scale model. This is canceled by an opposite
change in $\sigma_{\mathrm{ext}}^{2}$ by choosing excitatory and
inhibitory Poisson rates
\begin{equation}
r_{\mathrm{e,ext}}=r_{\mathrm{e,0}}+\frac{(1-\iota)\sigma_{\mathrm{int}}^{2}}{\tau_{\mathrm{m}}J_{\mathrm{ext}}^{2}\left(1+g\right)},\label{eq:r_e_ext}
\end{equation}
\begin{equation}
r_{\mathrm{i,ext}}=\frac{(1-\iota)\sigma_{\mathrm{int}}^{2}}{\tau_{\mathrm{m}}J_{\mathrm{ext}}^{2}g(1+g)},\label{eq:r_i_ext}
\end{equation}
where $r_{\mathrm{e,0}}$ is the Poisson rate in the full-scale
model, and the excitatory and inhibitory synapses have weights $J_{\mathrm{ext}}$
and $-g\, J_{\mathrm{ext}}$, respectively. Equations \eqref{eq:r_e_ext}
and \eqref{eq:r_i_ext} match eq. (E.1) in \cite{Helias13_023002}
except for the $1+g$ in the denominator, which was there erroneously
given as $1+g^{2}$. Since downscaling $K$ implies $\iota>1$, it
is seen that the required rate of the inhibitory inputs is negative.
Therefore, this method only allows upscaling. An alternative is to
use a balanced Poisson drive with weights $J_{\mathrm{ext}}$ and
$-\, J_{\mathrm{ext}}$, choosing the rate of both excitatory and
inhibitory inputs to generate the desired variance, and adding a DC
drive $I_{\mathrm{ext}}$ to recover the mean input,
\begin{eqnarray}
r_{\mathrm{e,ext}}=r_{\mathrm{i,ext}} & = & \frac{r_{\mathrm{e,0}}}{2}+\frac{(1-\iota)\sigma_{\mathrm{int}}^{2}}{2\tau_{\mathrm{m}}J_{\mathrm{ext}}^{2}},\label{eq:balanced_input_rate}\\
I_{\mathrm{ext}} & = & \tau_{\mathrm{m}}r_{\mathrm{e,0}}J_{\mathrm{ext}}.\label{eq:balanced_input_DC}
\end{eqnarray}
In this manner, the network can be downscaled up to the point where
the variance of the external drive vanishes. Substituting this condition
into \eqref{eq:generic_scaling}, the same expression for the minimal
in-degree scaling factor \eqref{eq:kappa_min} is obtained as for
the binary network.

\subsection{Robustness of correlation-preserving scaling\label{sub:Generalizability}}

\begin{figure*}
\centering{}\includegraphics{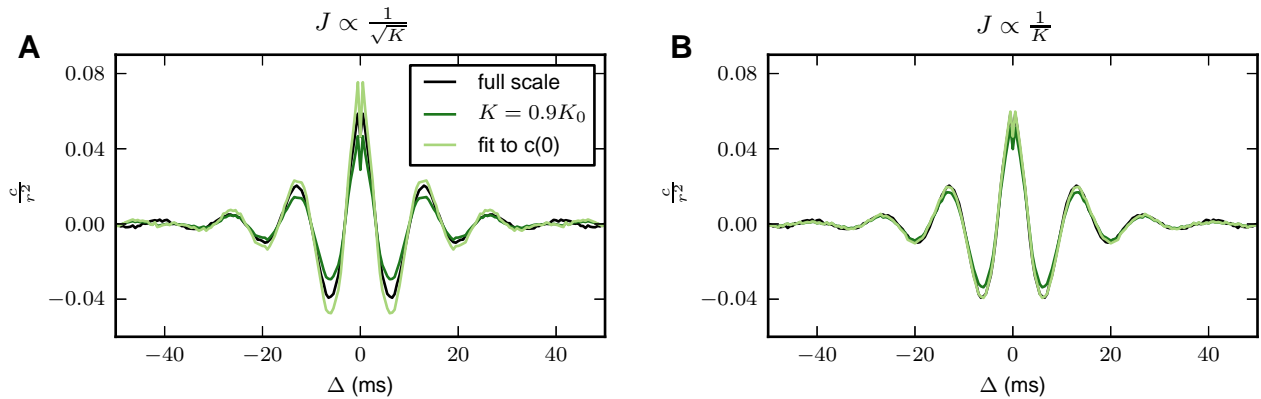}\protect\caption{Within the restrictive bounds imposed by \eqref{eq:kappa_min}, preserving
effective connectivity can preserve correlations also in a complex
network. Simulation results for the cortical microcircuit at full
scale and with in-degrees reduced to $90\%$. Synaptic strengths are
scaled as indicated, and the external drive is adjusted to restore
the working point. Mean pairwise cross-covariances are shown for population
2/3E. Qualitatively identical results are obtained within and across
other populations. The simulation duration is $30\,\mathrm{s}$ and
covariances are determined with a resolution of $0.5\,\mathrm{ms}$.
To enable downscaling with $J\propto1/K$, the excitatory Poisson
input of the original implementation of \cite{Potjans14_785} is replaced
by balanced inhibitory and excitatory Poisson input with a DC drive
according to \eqref{eq:balanced_input_rate} and \eqref{eq:balanced_input_DC}.
\textbf{A} Scaling synaptic strengths as\textbf{ }$J\propto1/\sqrt{K}$
changes the mean covariance. Light green curve: stretching the covariance
of the scaled network along the vertical axis to match the zero-lag
correlation of the full-scale network shows that not only the size
but also the temporal structure of the covariance is affected.\textbf{
B} Scaling synaptic strengths as $J\propto1/K$ closely preserves
the covariance of the full-scale network. However, note that this
scaling is only applicable down to the in-degree scaling factor given
by \eqref{eq:kappa_min}, which for this example is approximately
$0.9$. \label{fig:microcircuit_corr}}
\end{figure*}

In this section, we show that the scaling $J\propto1/K$, which maintains
the population-level feedback quantified by the effective connectivity,
can preserve correlations (within the bounds given in \nameref{sub:Limit-to-in-degree})
under fairly general conditions. To this end, we consider two types
of networks: 1. a multi-layer cortical microcircuit model with distributed
in- and out-degrees and lognormally distributed synaptic strengths
(cf. \nameref{sub:Network-structure-and-notation}); 2. a two-population
LIF network with different mean firing rates (parameters in \prettyref{tab:spiking_network-1}).
\begin{table}
\centering{}%
\begin{tabular}{|c|c|c|}
\hline 
number of excitatory neurons & $N$ & $8000$\tabularnewline
\hline 
relative inhibitory population size & $\gamma$ & $0.25$\tabularnewline
\hline 
membrane time constant & $\tau_{\mathrm{m}}$ & $20\,\mathrm{ms}$\tabularnewline
\hline 
synaptic time constant & $\tau_{\mathrm{s}}$ & $2\,\mathrm{ms}$\tabularnewline
\hline 
refractory period & $\tau_{\mathrm{ref}}$ & $2\,\mathrm{ms}$\tabularnewline
\hline 
membrane resistance & $R_{\mathrm{m}}$ & $20\,\mathrm{M\Omega}$\tabularnewline
\hline 
resting and reset potential & $V_{\mathrm{r}}$ & $0\,\mathrm{mV}$\tabularnewline
\hline 
threshold & $\theta$ & $15\,\mathrm{mV}$\tabularnewline
\hline 
connection probability & $p$ & $0.1$\tabularnewline
\hline 
transmission delay & $d$ & $3\,\mathrm{ms}$\tabularnewline
\hline 
excitatory synaptic weight & $J$ & $\mathrm{0.1\, mV}$\tabularnewline
\hline 
relative inhibitory synaptic weight & $g$ & $5$\tabularnewline
\hline 
mean and standard deviation of external drive & $(\mu_{\mathrm{ext}},\;\sigma_{\mathrm{ext}})$ & $(10\,\mathrm{mV},\;5\,\mathrm{mV})$; \tabularnewline
 &  & $(25\,\mathrm{mV},\;20\,\mathrm{mV})$\tabularnewline
\hline 
\end{tabular}\protect\caption{Full-scale parameters of the two-population spiking networks used
to demonstrate the robustness of $J\propto1/K$ scaling to mean firing
rates. The two networks are distinguished by their external drives.}
\label{tab:spiking_network-1}
\end{table}
For both types of models, we contrast the scaling $J\propto1/K$ with
$J\propto1/\sqrt{K}$, in each case maintaining the working point
given by \eqref{eq:mu} and \eqref{eq:sigma}. \prettyref{fig:microcircuit_corr}
illustrates that the former closely preserves average pairwise cross-covariances
in the cortical microcircuit model, whereas the latter changes both
their size and temporal structure.

\begin{figure}
\centering{}\includegraphics{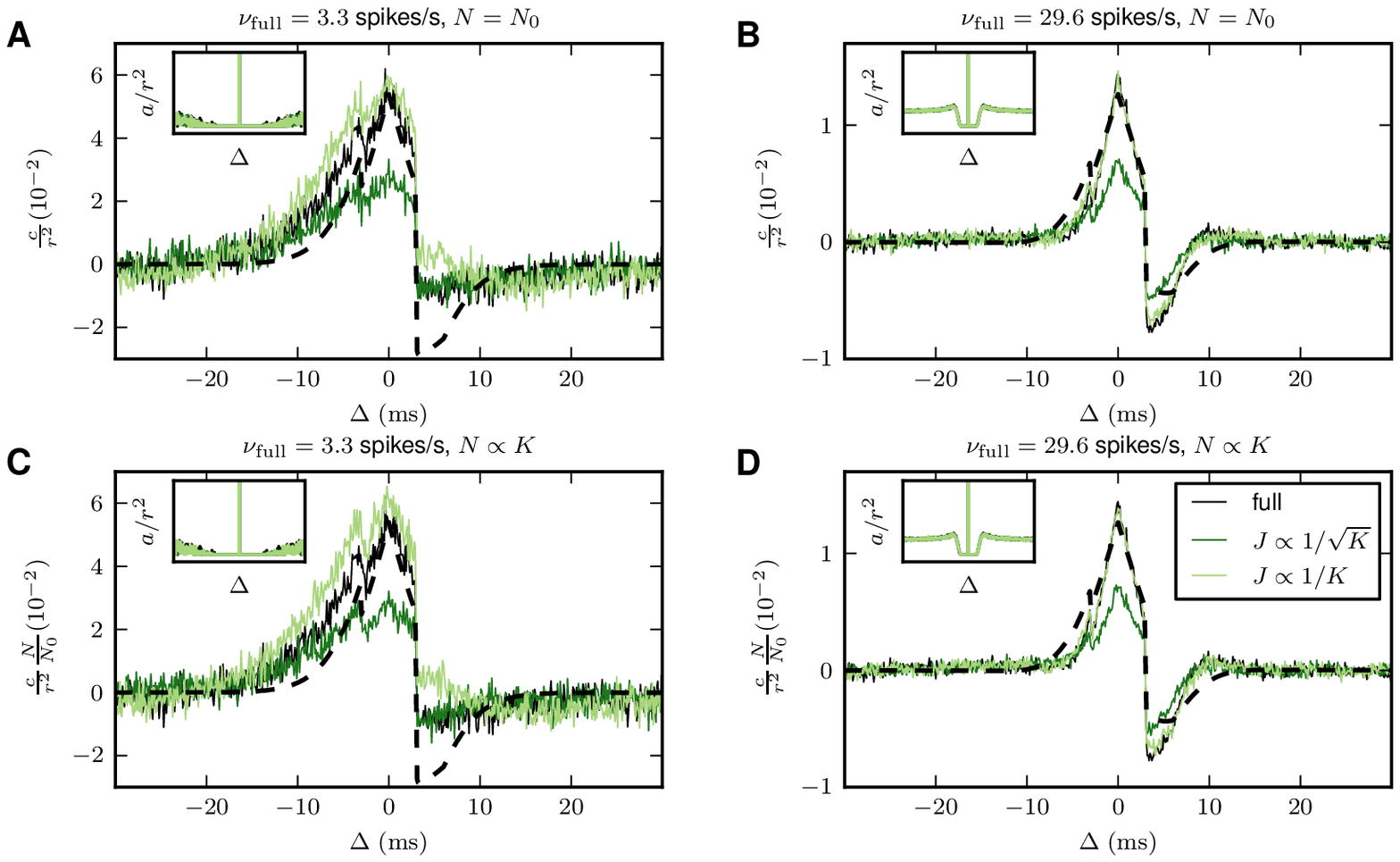}\protect\caption{Scaling synaptic strengths as $J\propto1/K$ can preserve correlations
in networks with widely different firing rates. Results of simulations
of a LIF network consisting of one excitatory and one inhibitory population
(\prettyref{tab:spiking_network-1}). Average cross-covariances are
determined with a resolution of $0.1\,\mathrm{ms}$ and are shown
for excitatory-inhibitory neuron pairs. Each network receives a balanced
Poisson drive with excitatory and inhibitory rates both given by $\sigma_{\mathrm{ext}}^{2}/(2\,\tau_{\mathrm{m}}\, J^{2})$,
where $\sigma_{\mathrm{ext}}^{2}$ is chosen to maintain the working
point of the full-scale network. The synaptic strengths for the external
drive are $0.1\,\mathrm{mV}$ and $-0.1\,\mathrm{mV}$ for excitatory
and inhibitory synapses, respectively. A DC drive with strength $\mu_{\mathrm{ext}}$
is similarly adjusted to maintain the full-scale working point. All
networks are simulated for $100\,\mathrm{s}$. For each population,
cross-covariances are computed as averages over all neuron pairs across
two disjoint groups of $\mathcal{N}\times1000$ neurons, where $\mathcal{N}$
is the scaling factor for the number of neurons (a given pair has
one neuron in each group). Autocovariances are computed as averages
over $100$ neurons in each population. \textbf{A}, \textbf{B }Reducing
in-degrees $K$ to $50\%$ while the number of neurons $N$ is held
constant, $J\propto1/K$ closely preserves both the size and the shape
of the covariances, while $J\propto1/\sqrt{K}$ diminishes their size.
\textbf{C},\textbf{ D} Reducing both $N$ and $K$ to $50\%$, covariance
sizes scale with $1/N$ for $J\propto1/K$ but with a different factor
for $J\propto1/\sqrt{K}$. Dashed curves represent theoretical predictions.
The insets show mean autocovariances for time lags $\Delta\in(-30,\,30)\:\mathrm{ms}$.
\label{fig:LIF_network_rate_comparison}}
\end{figure}
\prettyref{fig:LIF_network_rate_comparison} demonstrates the robustness
of $J\propto1/K$ scaling to the firing rate of the network. In this
example, both the full-scale network and the downscaled networks receive
a balanced Poisson drive producing the desired variance, while the
mean input is provided by a DC drive. By changing the parameters of
the external drive, we create two networks each with irregular spiking
but with widely different mean rates ($3.3$ spikes/s and $29.6$
spikes/s). Downscaling only the number of synapses but not the number
of neurons, both the temporal structure and the size of the correlations
are closely preserved. Reducing the in-degrees and the number of neurons
$N$ by the same factor, the correlations are scaled by $1/N$. Hence,
the correlations of the full-scale network of size $N_{0}$ can be
estimated simply by multiplying those of the reduced network by $N/N_{0}$.
In contrast, $J\propto1/\sqrt{K}$ changes correlation sizes even
when $N$ is held constant, and combined scaling of $N$ and $K$
can therefore not simply be compensated for by the factor $N/N_{0}$.
In the high-rate network, the spiking statistics of the neurons is
non-Poissonian, as seen from the gap in the autocorrelations (insets
in \prettyref{fig:LIF_network_rate_comparison}B, D). Nevertheless,
$J\propto1/K$ preserves the correlations more closely than $J\propto1/\sqrt{K}$,
showing that the predicted scaling properties hold beyond the strict
domain of validity of the underlying theory.

\subsection{Zero-lag correlations in binary network\label{sub:Zero-lag-corr}}

\begin{figure*}
\centering{}\includegraphics{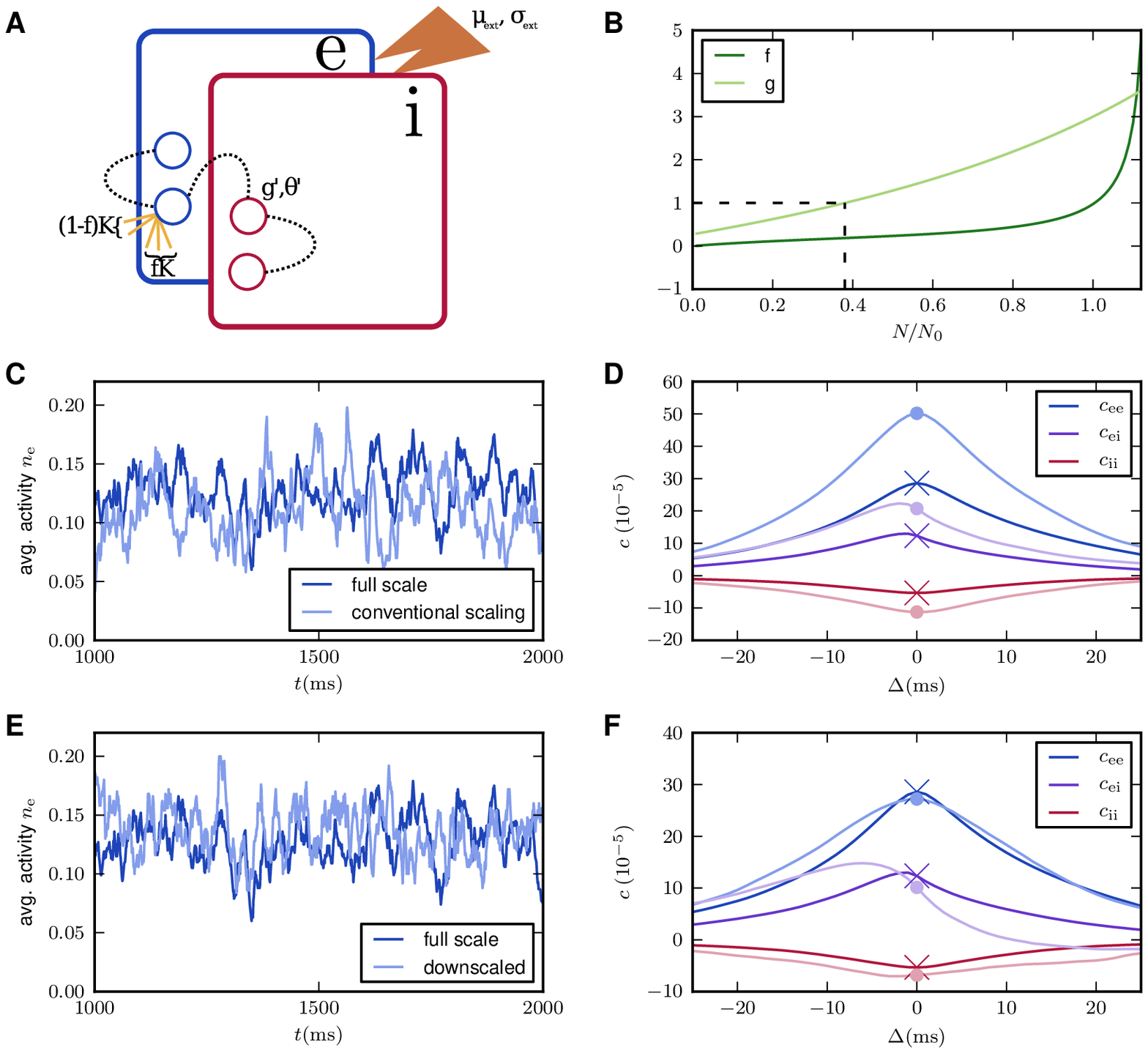}\protect\caption{Binary network scaling that approximately preserves both mean activities
and zero-lag covariances. \textbf{A} Increased covariances due to
reduced network size can be countered by a change in the relative
inhibitory synaptic weight combined with a redistribution of the synapses
so that a fraction comes from outside the network. Adjusting a combination
of the threshold and external drive restores the working point. \textbf{B}
Scaling parameters versus relative network size for an example network.
Since $\gamma=1$ in this example, the scaling only works down to
$g=1$ (indicated by the horizontal and vertical dashed lines): Lower
values of $g$ only allow a silent or fully active network as steady
states. \textbf{C},\textbf{ E} The mean activities are well preserved
both by the conventional scaling in \eqref{eq:Vreeswijk_scaling}
with an appropriate adjustment of $\theta$ (panel \textbf{C}), and
by the method proposed here (panel \textbf{E}). \textbf{D},\textbf{
F} Conventional scaling increases the magnitude of zero-lag covariances
in simulated data (panel \textbf{D}), while the proposed method preserves
them (panel \textbf{F}). Dark colors: full-scale network. Light colors:
downscaled network. Crosses and dots indicate zero-lag correlations
in the full-scale and downscaled networks, respectively.}
\label{fig:scaling_binary_network}
\end{figure*}
\begin{table}
\centering{}%
\begin{tabular}{|c|c|c|}
\hline 
number of excitatory neurons & $N$ & $1000$\tabularnewline
\hline 
relative inhibitory population size & $\gamma$ & $1$\tabularnewline
\hline 
neuron time constant & $\tau$ & $10\,\mathrm{ms}$\tabularnewline
\hline 
threshold & $\theta$ & $-3$\tabularnewline
\hline 
connection probability & $p$ & $0.2$\tabularnewline
\hline 
transmission delay & $d$ & $0.1\,\mathrm{ms}$\tabularnewline
\hline 
excitatory synaptic weight & $J$ & $1/\sqrt{1000}$\tabularnewline
\hline 
relative inhibitory synaptic weight & $g$ & $3$\tabularnewline
\hline 
external drive & $m_{\mathrm{x}}$ & $0$\tabularnewline
\hline 
\end{tabular}\protect\caption{Parameters of the symmetric binary network.}
\label{tab:binary_network}
\end{table}

Although it is not generally possible to keep mean activities and
correlations invariant upon downscaling, transformations may be found
when only one aspect of the correlations is important, such as their
zero-lag values. We illustrate this using a simple, randomly connected
binary network of $N$ excitatory and $\gamma N$ inhibitory binary
neurons, where each neuron receives $K=pN$ excitatory and $\gamma K$
inhibitory inputs. The parameters are given in \prettyref{tab:binary_network}.
The linearized effective connectivity matrix for this example is 
\begin{equation}
\mathbf{W}=S(\mu,\sigma)JK\begin{pmatrix}1 & -\gamma g\\
1 & -\gamma g
\end{pmatrix}.
\end{equation}

When the threshold $\theta$ is $\leq0$, the network is spontaneously
active without external inputs. In the diffusion approximation and
assuming stationarity, the mean zero-lag cross-covariances between
pairs of neurons from each population can be estimated from \eqref{eq:corr_self_consistent_pop}
(see also \cite{Grytskyy13_258})

\[
\left[\begin{pmatrix}1 & 0\\
0 & 1
\end{pmatrix}-\frac{W_{\mathrm{e}}}{2}\begin{pmatrix}2-\gamma g & -\gamma g\\
1 & 1-2\gamma g
\end{pmatrix}\right]\begin{pmatrix}c_{\mathrm{ee}}\\
c_{\mathrm{ii}}
\end{pmatrix}=\frac{W_{\mathrm{e}}a}{N}\begin{pmatrix}1\\
-g
\end{pmatrix}
\]
\begin{equation}
c_{\mathrm{ei}}=c_{\mathrm{ie}}=\frac{1}{2}(c_{\mathrm{ee}}+c_{\mathrm{ii}}),\label{eq:binary_corr}
\end{equation}
where the subscripts $\mathrm{e}$ and $\mathrm{i}$ respectively
denote excitatory and inhibitory populations. Moreover, $W_{\mathrm{e}}$
is the effective excitatory coupling,
\begin{equation}
W_{\mathrm{e}}=S(\mu,\sigma)JK.
\end{equation}
with $S$ the susceptibility as defined in \eqref{eq:susceptibility}.
Furthermore, $a$ is the variance of the single-neuron activity,
\begin{equation}
a=\left\langle n\right\rangle (1-\left\langle n\right\rangle ),
\end{equation}
which is identical for the excitatory and inhibitory populations.
The mean input to each neuron is given by {[}cf. \eqref{eq:mu}{]},
\begin{equation}
\mu=JK(1-\gamma g)\left\langle n\right\rangle ,
\end{equation}
and, under the assumption of near-independence of the neurons, the
variance of the inputs is well approximated by the sum of the variances
from each sending neuron {[}cf. \eqref{eq:sigma}{]},
\begin{equation}
\sigma^{2}=J^{2}K(1+\gamma g^{2})\left\langle n\right\rangle (1-\left\langle n\right\rangle ).
\end{equation}
Finally, the mean activity can be obtained from the self-consistency
relation \eqref{eq:binary_mean_activity}.

Equation \eqref{eq:binary_corr} shows that, when excitatory and inhibitory
synaptic weights are scaled equally, the covariances scale with $1/N$
as long as the network feedback is strong $(W_{\mathrm{e}}\gg1),$
(for this argument, we assume that $\left\langle n\right\rangle $
is held constant, which may be achieved by adjusting a combination
of $\theta$ and the external drive). Hence, conventional downscaling
of population sizes tends to increase covariances.

We use \eqref{eq:binary_corr} to perform a more sophisticated downscaling
(cf. \prettyref{fig:scaling_binary_network}). Let the new size of
the excitatory population be $N^{\prime}$. Equation \eqref{eq:binary_corr}
shows that the covariances can only be preserved when a combination
of $W_{\mathrm{e}}$, $\gamma$, and $g$ is adjusted. We take $\gamma$
constant, and apply the transformation 
\begin{equation}
W_{\mathrm{e}}\rightarrow fW_{\mathrm{e}};\; g\rightarrow g^{\prime}.
\end{equation}
Solving \eqref{eq:binary_corr} for $f$ and $g^{\prime}$ yields
(cf. \prettyref{fig:scaling_binary_network}B)
\begin{equation}
f=\frac{\frac{a\, c_{\mathrm{ee}}}{N^{\prime}}+\frac{\gamma\, c_{\mathrm{ii}}}{2}(c_{\mathrm{ee}}-c_{\mathrm{ii}})}{W_{\mathrm{e}}\left[\left(\frac{a}{N^{\prime}}+c_{\mathrm{ee}}\right)\left(\frac{a}{N}+\gamma c_{\mathrm{ii}}\right)-\frac{\gamma}{4}(c_{\mathrm{ee}}+c_{\mathrm{ii}})^{2}\right]}
\end{equation}
\begin{equation}
g^{\prime}=\frac{c_{\mathrm{ee}}(c_{\mathrm{ee}}-c_{\mathrm{ii}})-\frac{2a}{N^{\prime}}c_{\mathrm{ii}}}{\gamma\, c_{\mathrm{ii}}(c_{\mathrm{ee}}-c_{\mathrm{ii}})+\frac{2a}{N^{\prime}}c_{\mathrm{ee}}}.\label{eq:g_new}
\end{equation}
The change in $W_{\mathrm{e}}$ can be captured by $K\rightarrow f\, K$
as long as the working point $(\mu,\,\sigma$) is maintained. This
intuitively corresponds to a redistribution of the synapses so that
a fraction $f$ comes from inside the network, and $1-f$ from outside
(cf. \prettyref{fig:scaling_binary_network}A). However, the external
drive does not have the same mean and variance as the internal inputs,
since it needs to make up for the change in $g.$ The external input
can be modeled as a Gaussian noise with parameters 
\begin{equation}
\mu_{\mathrm{ext}}=KJ(1-\gamma g)\left\langle n\right\rangle -fKJ(1-\gamma g^{\prime})\left\langle n\right\rangle 
\end{equation}
\begin{equation}
\sigma_{\mathrm{ext}}^{2}=KJ^{2}(1+\gamma g^{2})a-fKJ^{2}(1+\gamma g^{\prime2})a,
\end{equation}
independent for each neuron. 

An alternative is to perform the downscaling in two steps: First change
the relative inhibitory weights according to \eqref{eq:g_new} but
keep the connection probability constant. The mean activity can be
preserved by solving \eqref{eq:binary_mean_activity} for $\theta$,
but the covariances are changed. The second step, which restores the
original covariances, then amounts to redistributing the synapses
so that a fraction $\tilde{f}$ comes from inside the network, and
$1-\tilde{f}$ from outside, where the external (non-modeled) neurons
have the same mean activity as those inside the network. This mean
activity is negative, as the balanced regime implies stronger inhibition
than excitation. Note that $\tilde{f}\neq f$, since $W_{\mathrm{e}}$
changes already in the first step.

The requirement that inhibition dominate excitation places a lower
limit on the network size for which the scaling is effective. The
reason is that $g$ decreases with network size, so that a bifurcation
occurs at $g=1/\gamma$, beyond which the only steady states correspond
to a silent network or a fully active one.

\subsection{Symmetric two-population spiking network\label{sub:Symmetric-two-population-spiking}}

We have seen that the one-to-one relationship between effective connectivity
and correlations does not hold in certain degenerate cases. Here we
consider such a degenerate case and perform a scaling that preserves
mean activities as well as both the size and the temporal structure
of the correlations under reductions in both the number of neurons
and the number of synapses. The network consists of one excitatory
and one inhibitory population of LIF neurons with a population-independent
connection probability and vanishing transmission delays. Due to the
appearance of the eigenvalues in the numerator of the expression for
the correlations in LIF networks {[}cf.  \eqref{eq:LIF_corr_Fourier_transform}
and \eqref{eq:LIF_cov_matrix}{]}, such networks are subject to a
reduced number of constraints when $\mathbf{W}$ has a zero eigenvalue,
as this leaves a freedom to change the corresponding eigenvectors.
Furthermore, identically vanishing delays greatly simplify the equations
for the covariances.

The single-neuron and network parameters are as in \prettyref{tab:spiking_network-1}
except that, here, $N=10,000$, $J=0.2\,\mathrm{mV}$, and the external
drive is chosen such that the mean and standard deviation of the total
input to each neuron are $\mu=15\,\mathrm{mV}$, $\sigma=10\,\mathrm{mV}$.
Furthermore, the delay is chosen equal to the simulation time step
to approximate $d=0$, which we assume here. The effective connectivity
matrix for this network is 
\begin{equation}
\mathbf{W}=w\, K\,\begin{pmatrix}1 & -\gamma g\\
1 & -\gamma g
\end{pmatrix},\label{eq:LIF_W}
\end{equation}
where $w=\partial r_{\mathrm{target}}/\partial r_{\mathrm{source}}$
is the effective excitatory synaptic weight obtained as the derivative
of \eqref{eq:siegert}. Here, we take into account the dependence
of $w$ on $J$ to quadratic order. The inhibitory weight is approximated
as $gw$ to allow an analytical expression for the relative inhibitory
weight in the scaled network to be derived. The left and right eigenvectors
are $\mathbf{v}^{1}=\frac{1}{\sqrt{1-\gamma g}}\,\begin{pmatrix}1\\
-\gamma g
\end{pmatrix},$ $\mathbf{u}^{1}=\frac{1}{\sqrt{1-\gamma g}}\,\begin{pmatrix}1\\
1
\end{pmatrix}$ corresponding to eigenvalue $L=w\, K\left(1-\gamma\, g\right)$ and
$\mathbf{v}^{2}=\frac{1}{\sqrt{1-\frac{1}{\gamma g}}}\,\begin{pmatrix}1\\
-1
\end{pmatrix},$ $\mathbf{u}^{2}=\frac{1}{\sqrt{1-\frac{1}{\gamma g}}}\,\begin{pmatrix}1\\
\frac{1}{\gamma g}
\end{pmatrix}$ corresponding to eigenvalue $0$. The normalization is chosen such
that the bi-orthogonality condition \eqref{eq:biorthogonality} is
fulfilled. 

A transformed connectivity matrix should have the same eigenvalues
as $\mathbf{W}$, and can thus be written as
\begin{eqnarray}
\mathbf{W}^{\prime} & = & w^{\prime}K^{\prime}\begin{pmatrix}1 & -b\\
c & -bc
\end{pmatrix}\\
\mathrm{where}\quad b & = & \frac{1}{c}\left[1-\frac{wK}{w^{\prime}K^{\prime}}(1-\gamma g)\right].
\end{eqnarray}
Denote the new population sizes by $N_{1}$ and $N_{2}$. Equating
the covariances before and after the transformation yields using \eqref{eq:LIF_cov_matrix}
and $A^{jk}=\mathbf{v}^{jT}\mathbf{A}\mathbf{v}^{k}$ {[}cf. \eqref{eq:c_ik}{]},
\begin{multline}
\frac{\frac{a_{1}}{N}+\gamma g^{2}\frac{a_{2}}{N}}{(1-\gamma g)^{2}(2-2L)}\begin{pmatrix}1 & 1\\
1 & 1
\end{pmatrix}+\frac{\frac{a_{1}}{N}+g\frac{a_{2}}{N}}{(2-\gamma g-\frac{1}{\gamma g})(2-L)}\begin{pmatrix}1 & \frac{1}{\gamma g}\\
1 & \frac{1}{\gamma g}
\end{pmatrix}\\
=\frac{\frac{a_{1}}{N_{1}}+b^{2}\frac{a_{2}}{N_{2}}}{(1-bc)^{2}(2-2L)}\begin{pmatrix}1 & c\\
c & c^{2}
\end{pmatrix}+\frac{\frac{a_{1}}{N_{1}}+\frac{b}{c}\frac{a_{2}}{N_{2}}}{(2-bc-\frac{1}{bc})(2-L)}\begin{pmatrix}1 & \frac{1}{b}\\
c & \frac{c}{b}
\end{pmatrix}.\label{eq:equating_LIF_cov}
\end{multline}
In \eqref{eq:equating_LIF_cov} we have assumed that the working
points, and thus $a_{1}$ and $a_{2}$, are preserved, which may be
achieved with an appropriate external drive as long as the corresponding
variance remains positive. The four equations are simultaneously
solved by
\begin{eqnarray}
N_{1} & = & \frac{Nw^{\prime}K^{\prime}a_{1}\left(2-L\right)}{wKga_{2}\left(wK-w^{\prime}K^{\prime}\right)+a_{1}\left[2w^{\prime}K^{\prime}-wK(w^{\prime}K^{\prime}-\gamma gwK)\right]}\nonumber \\
N_{2} & = & \frac{Na_{2}}{wK}\frac{L(L-w^{\prime}K^{\prime}-2)+2w^{\prime}K^{\prime}}{ga_{2}\left(2-L\right)+(w^{\prime}K^{\prime}-wK)(a_{1}+ga_{2})}\nonumber \\
c & = & 1,\label{eq:N1_N2_c}
\end{eqnarray}
where $K^{\prime}w^{\prime}$ may be chosen freely. Thus, the new
connectivity matrix reads
\begin{equation}
\mathbf{W}^{\prime}=w^{\prime}\, K^{\prime}\begin{pmatrix}1 & \frac{wK}{w^{\prime}K^{\prime}}(1-\gamma\, g)-1\\
1 & \frac{wK}{w^{\prime}K^{\prime}}(1-\gamma\, g)-1
\end{pmatrix},
\end{equation}
which may also be cast into the form
\begin{equation}
\mathbf{W}^{\prime}=w^{\prime}\, K^{\prime}\begin{pmatrix}1 & -\gamma^{\prime}\, g^{\prime}\\
1 & -\gamma^{\prime}\, g^{\prime}
\end{pmatrix},
\end{equation}
where $\gamma^{\prime}=N_{2}/N_{1}$ and $g^{\prime}=\frac{w^{\prime}K^{\prime}-L}{w^{\prime}K^{\prime}\gamma^{\prime}}$.
\begin{figure*}
\centering{}\includegraphics{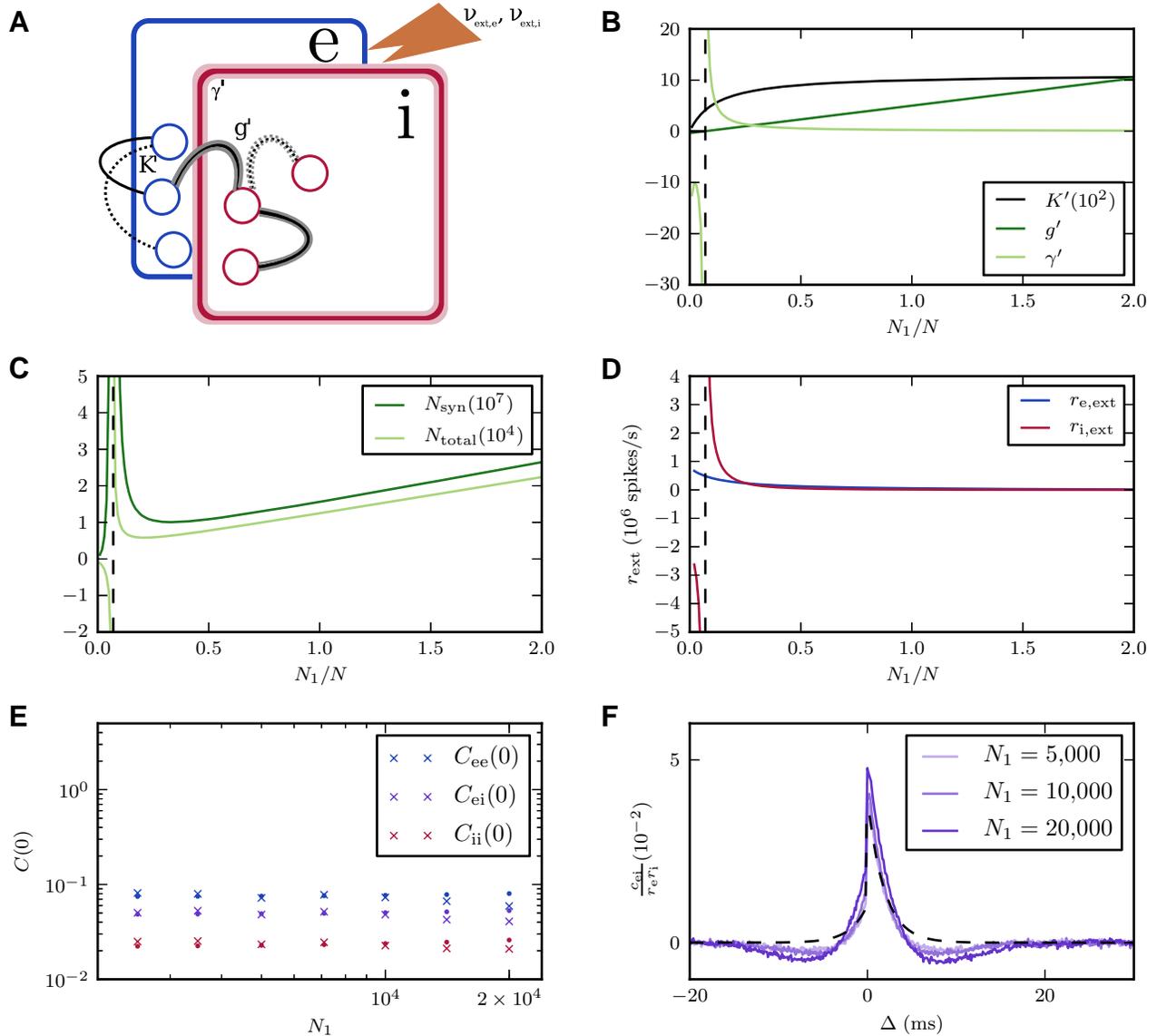}\protect\caption{Spiking network scaling that approximately preserves mean firing rates
and covariances. \textbf{A} Diagram illustrating the network and indicating
the parameters that are adjusted. \textbf{B} Excitatory in-degrees
$K^{\prime}$, relative inhibitory synaptic weight $g^{\prime}$,
and relative number of inhibitory neurons $\gamma^{\prime}$ versus
scaling factor $N_{1}/N$. The dashed vertical line indicates the
limit below which the scaling fails. \textbf{C} Total number of neurons
$N_{\mathrm{total}}=(1+\gamma^{\prime})N_{1}$ and total number of
synapses $N_{\mathrm{syn}}=(1+\gamma^{\prime})^{2}K^{\prime}N_{1}$
versus scaling factor. \textbf{D} Rates of external excitatory and
inhibitory Poisson inputs necessary for keeping firing rates constant.
Average firing rates are between $23.1$ and $23.5$ spikes/s for
both excitatory and inhibitory populations and all network sizes.
\textbf{E} Integrated covariances, corresponding to zero-frequency
components in the Fourier domain. Crosses: simulation results, dots:
theoretical predictions. \textbf{F} Average covariance between excitatory-inhibitory
neuron pairs for different network sizes. The dashed curve indicates
the theoretical prediction for $N=10,000$. Each network was simulated
for $100\,\mathrm{s}$.}
\label{fig:simple_spiking_network}
\end{figure*}

When the populations receive statistically identical external inputs,
we have $a_{1}=a_{2}=r$, since the internal inputs are also equal.
\prettyref{fig:simple_spiking_network} illustrates the network scaling
for the choice $w^{\prime}=w$. Results are shown as a function of
the relative size $N_{1}/N$ of the excitatory population. External
drive is provided at each network size to keep the mean and standard
deviation of the total inputs to each neuron at the level indicated.
The mean is supplied as a constant current input, while the variability
is afforded by Poisson inputs according to \eqref{eq:r_e_ext} and
\eqref{eq:r_i_ext} (\prettyref{fig:simple_spiking_network}D). It
is seen that the transformations (\prettyref{fig:simple_spiking_network}B)
are able to reduce both the total numbers of neurons and the total
number of synapses (\prettyref{fig:simple_spiking_network}C) while
approximately preserving covariance sizes and shapes (\prettyref{fig:simple_spiking_network}E,F).
Small fluctuations in the theoretical predictions in \prettyref{fig:simple_spiking_network}E
are due to the discreteness of numbers of neurons and synapses, and
deviations of the effective inhibitory weight from the linear approximation
$g\, w$. The fact that the theoretical prediction in \prettyref{fig:simple_spiking_network}F
misses the small dips around $t=0$ may be due to the approximation
of the autocorrelations by delta functions, eliminating the relative
refractoriness due to the reset. The numbers of neurons and synapses
increase again below some $N_{1}/N$, and diverge as $g^{\prime}$
becomes zero. This limits the scalability despite the additional freedom
provided by the symmetry.

\section{Discussion}

By applying and extending the theory of correlations in asynchronous
networks of binary and networks of leaky integrate-and-fire (LIF)
neurons, our present work shows that the scalability of numbers of
neurons and synapses is fundamentally limited if mean activities and
pairwise averaged activity correlations are to be preserved. We analytically
derive a limit on the reducibility of the number of incoming synapses
per neuron, $K$ (the in-degree), which depends on the variance of
the external drive, and which indirectly restricts the scalability
of the number of neurons. Within these restrictive bounds, we propose
a scaling of the synaptic strengths $J$ and the external drive with
$K$ that can preserve mean activities and the size and temporal structure
of pairwise averaged correlations. Mean activities can be approximately
preserved by maintaining the mean and variance of the total input
currents to the neurons, also referred to as the \textit{working point}.
The temporal structure of pairwise averaged correlations depends on
the \textit{effective connectivity}, a measure of the effective influence
of source populations on target populations determined both by the
physical connectivity and the working point of the target neurons.
When the dependence of the effective connectivity on the synaptic
strengths $J$ is linearized, it can be written as $SJK$, where $S$
is the susceptibility of the target neurons (quantifying the change
in output activity for a unit change in input). Scalings and analytical
predictions of pairwise averaged correlations are tested using direct
simulations of randomly connected networks of excitatory and inhibitory
neurons.

Our most important findings are:
\begin{enumerate}
\item[i)]  The population-level effective connectivity matrix and pairwise
averaged correlations are linked by a one-to-one mapping except in
degenerate cases. Therefore, with few exceptions, any network scaling
that preserves the correlations needs to preserve the effective connectivity.

\item[ii)]  The most straightforward way of simultaneously preserving mean activities
and pairwise averaged correlations is to change the synaptic strengths
in inverse proportion to the in-degrees ($J\propto1/K$), and to adjust
the variance of the external drive to make up for the change in variance
of inputs from within the network. Other scalings, such as $J\propto1/\sqrt{K}$,
can in principle also preserve both mean activities and pairwise averaged
correlations, but then change the working point (hence the neuronal
susceptibility determining the strength of stimulus responses, and
the degree to which the activity is mean- or fluctuation-driven),
and are analytically intractable for LIF networks due to the complicated
dependence of the firing rates and the impulse response on the mean
and variance of the inputs. 
\item[iii)]  When downscaling the in-degrees $K$ and scaling synaptic strengths
as $J\propto1/K$, the variance of inputs from within the network
increases, so that the variance of external inputs needs to be decreased
to restore the working point. This is only possible up to the point
where the variance of the external drive vanishes. The minimal in-degree
scaling factor equals the ratio between the variance of inputs coming
from within the network, and the total input variance due to both
internal inputs and the external drive. The same limit to in-degree
scaling holds more generally for scalings that simultaneously preserve
mean activities and correlations. Thus, in the absence of a variable
external drive, no downscaling is possible without changing mean activities,
correlations, or both.
\item[iv)]  Within the identified restrictive bounds, the scaling $J\propto1/K$,
where the external variance is adjusted to maintain the working point,
can preserve mean activities and pairwise averaged correlations also
in asynchronous networks deviating from the assumptions of the analytical
theory presented here. We show this robustness for an example network
with distributed in- and out-degrees and distributed synaptic weights,
and for a network with non-Poissonian spiking.
\item[v)]  For a sufficiently large change in in-degrees, a scaling that affects
correlations can push the network from the linearly stable to an oscillatory
regime or vice versa.
\item[vi)]  Transformations derived using the diffusion approximation are able
to closely preserve the relevant quantities (mean activities, correlation
shapes and sizes) in simulated networks of binary and spiking neurons
within the given bounds. Reducing the number of neurons only increases
correlation magnitudes without affecting their structure in this approximation.However,
strong deviations from the assumptions of the diffusion approximation
can cause also correlation structure to change in simulated networks
under scalings originally constructed to maintain correlation structure.
This occurs for instance when a drastic reduction in network size
is coupled with a less than proportional reduction in in-degrees,
leading to large numbers of common inputs and increased synchrony.
Thus, the scalability of the number of neurons with available analytical
results is indirectly limited by the minimal in-degree scaling factor. 
\end{enumerate}
In conclusion, we have identified limits to the reducibility of neural
networks, even when only considering first- and second-order statistical
properties. Networks are inevitably irreducible in some sense, in
that downscaled networks are clearly not identical to their full-scale
counterparts. However, mean activity, a first-order macroscopic quantity,
can usually be preserved. The present work makes it clear that non-reducibility
already sets in at the second-order macroscopic level of correlations.
This does not imply a general minimal size for network models to be
valid, merely that each network in question needs to be studied near
its natural size to verify results from any scaled versions. 

Our analytical theory is based on the diffusion approximation, in
which inputs are treated as Gaussian noise, valid in the asynchronous
irregular regime when activities are sufficiently high and synaptic
weights are small. Moreover, external inputs are taken to be independent
across populations, and delays and time constants are assumed to be
unchanged under scaling. A further assumption of the theory is that
the dynamics is stationary and linearly stable. 

The one-to-one correspondence between effective connectivity and correlations
applies with a few exceptions. For non-identical populations with
different impulse responses, an analysis in the frequency domain demonstrates
the equivalence under the assumption that the correlation matrix is
invertible. An argument that assumes a diagonalizable effective connectivity
matrix extends the equivalence to identical populations apart from
cases where the effective connectivity matrix has eigenvalues that
are zero or degenerate. 

The equivalence of correlations and effective connectivity ties in
with efforts to infer structure from activity, not only in neuroscience
\cite{Kaminski01_145,Friston_03,Nykamp07_204,Timme_07_224101,Roudi11,Pernice12,Grytskyy13_NWG,Robinson14_012707}
but also in other disciplines \cite{Dhaeseleer00_707,Steuer03_1019,Psorakis12},
as it implies that one should in principle be able to find the only---and
therefore the real---effective connectivity that accounts for the
correlations. \textcolor{black}{Within the same framework as that
used here, \cite{Grytskyy13_NWG} show that knowledge of the cross-spectrum
at two distinct frequencies allows a unique reconstruction of the
effective connectivity matrix by splitting the covariance matrix into
symmetric and antisymmetric parts. The derivation considers a class
of transfer functions }\textcolor{black}{(the Fourier transform of
the neuronal impulse response) rather than any specific form, but
the transfer function is taken to be unique, whereas the present work
allows for differences between populations. Furthermore, we here present
a more straightforward derivation of the equivalence, not focused
on the practical aim of network reconstruction, and clarify the conditions
under which reconstruction is possible.}

In practice, using our results to infer structure from correlations
may not be straightforward, due to both deviations from the assumptions
of the theory and problems with measuring the relevant quantities.
\textcolor{black}{} For instance, neural activity is often nonstationary
\cite{Tyrcha13}, transfer functions are normally not measured directly,
and correlations are imperfectly known due to measurement noise. Furthermore,
inference of anatomical from functional connectivity (correlations)
is often done based on functional magnetic resonance imaging (fMRI)
measurements, which are sensitive only to very low frequencies and
therefore only allow the symmetric part of the effective connectivity
to be reliably determined \cite{Robinson14_012707}. The presence
of unobserved populations providing correlated input to two or more
observed populations can also hinder inference of network structure.
Thus, high-resolution measurements (e.g., two-photon microscopy combined
with optogenetics to record activity in a cell-type-specific manner
\cite{Helmchen09,Akemann13}) of networks with controlled input (e.g.,
in brain slices) hold the most promise for network reconstruction
from correlations.

The effects on correlation-based synaptic plasticity of scaling-related
changes in correlations may be partly compensated for by adjusting
the learning parameters. For instance, an increase in average correlation
size with factor $1/N$ without a change in temporal shape may be
to some extent countered by reducing the learning rate by the same
factor. Changes in the temporal structure of the correlations are
more difficult to compensate for. When learning is linear or slow,
so that the learning function can be approximated as constant (independent
of the weights), the mean drift in the synaptic weights is determined
by the integral of the product of the correlations and the learning
function \cite{Kempter99,Kunkel11_00160}. Therefore, this mean drift
may be kept constant under a change in correlation shapes by adjusting
the learning function such that this product is preserved for all
time lags. However, given that the expression for the correlations
is a complicated function of the network parameters, the required
adjustment of the learning function will also be complex. Moreover,
the effects of this adjustment on precise patterns of weights are
difficult to predict, since the distribution of correlations between
neurons pairs may change under the proposed scalings, and this solution
does not apply when learning is fast and weight-dependent. 

The groundbreaking work of \cite{VanVreeswijk98_1321} identified
a dynamic balance between excitation and inhibition as a mechanism
for the asynchronous irregular activity in cortex, and showed that
$J\propto1/\sqrt{K}$ can robustly lead to a balanced state in the
limit $N\to\infty$ for constant $K/N$. However, it is not necessary
to scale synaptic weights as $1/\sqrt{K}$ in order to obtain a balanced
network state, even in the limit of infinite network size (and infinite
$K$). For instance, $J\propto1/K$ can retain balance in the infinite
size limit in the sense that the sum of the excitatory and inhibitory
inputs is small compared to each of these inputs separately. To retain
irregular activity with this scaling one merely needs to ensure a
variable external drive, as the internal variance vanishes for $N\to\infty$.
Moreover, in binary networks with neurons that have a Heaviside gain
function (a hard threshold) identical across neurons, one does not
even need a variable drive in order to stay in a balanced state \cite[p. 1360]{VanVreeswijk98_1321}.
This can be seen from a simple example of a network of $N$ excitatory
and $\gamma N$ inhibitory neurons with random connectivity with probability
$p$, where $J=J_{0}/N>0$ is the synaptic amplitude of an excitatory
synapse, and$-gJ$ the amplitude of an inhibitory synapse. The network
may receive a DC drive, which we absorb into the threshold $\theta$.
The summed input to each cell is then $\mu=pNJ(1-\gamma g)\, n$,
where $n\in[0,1]$ is the mean activity in the network. For a balanced
state to arise the negative feedback must be sufficiently strong,
so that the mean activity $n$ settles on a level where the summed
input is close to the threshold $\mu\simeq\theta$. This will always
be achieved if $pJ_{0}(1-\gamma g)<\theta<0$: in a completely activated
network ($n=1$) the summed input is below threshold, in a silent
network ($n=0$), the summed input is above threshold, hence the activity
will settle close to the value $n\simeq\theta/[pJ_{0}(1-\gamma g)]$.
As the variance of the synaptic input decreases with network size,
the latter estimate of the mean activity will become exact in the
limit $N\to\infty$. The underlying reason for both $1/K$ and $1/\sqrt{K}$
scaling to lead to a qualitatively identical balanced state is the
absence of a characteristic scale on which to measure the synaptic
input: the threshold is hard. Only by introducing a characteristic
scale, for example distributed values for the thresholds, the $1/K$
scaling with a DC drive will in the large $N$ limit lead to a freezing
of the balanced state due to the vanishing variance of the summed
input, while with either $1/\sqrt{K}$ scaling, or $1/K$ scaling
with a fluctuating external drive, the balanced state is conserved.

In \cite{VanVreeswijk98_1321}, $J\propto1/\sqrt{K}$ refers not only
to a comparison between differently-sized networks, but also to the
assumption that approximately $\sqrt{K}$ excitatory synapses need
to be active to reach spike threshold. However, this is also not a
necessary condition for balance, which can arise for a wide range
of synaptic strengths relative to threshold, as long as inhibition
is sufficiently strong compared to excitation. As discussed in \nameref{sub:Correlation-preserving-scaling},
with appropriately chosen external drive, $J$ even drops out of the
mean-field theory for binary networks with a Heaviside gain function
altogether \cite{Grytskyy13_258}. The difficulty in the interpretation
of the \cite{VanVreeswijk98_1321} results illustrates a more general
point: The primary goal of scaling studies is to identify the mechanisms
governing network dynamics. Nevertheless, these studies usually also
specify requirements on the robustness of the mechanism, leading to
scaling laws for network parameters that may be more restrictive
than a description of the mechanism per se. An example is the robustness
to strong synapses, defined such that activation of $\sim\sqrt{K}$
excitatory synapses suffices to reach threshold in the absence of
an external drive \cite[p. 1324]{VanVreeswijk98_1321}. This scenario
was considered in order to create a condition under which dynamic
balance is clearly \textit{necessary} for achieving asynchronous irregular
activity in balanced random networks, since combined inputs would
otherwise far exceed the threshold. However, dynamic balance can arise
also with weak synapses, e.g., with strength $\sim1/K$ of the distance
to threshold. Without questioning the value of scaling studies, which
can distill essential mechanisms and are sometimes possible where
finite-size analytical descriptions are intractable, this shows that
scaling laws need to be interpreted with care.

The issue of the interrelation between network size, synaptic strengths,
numbers of synapses per neuron, and activity is embedded in the wider
context of anatomical and physiological scaling laws observed experimentally.
In homeostatic synaptic plasticity, synaptic strengths are adjusted
in a manner that keeps the activity of the postsynaptic neurons within
a certain operating range \cite{Turrigiano08_422,Turrigiano98,Burrone2003_560}.
Since postsynaptic activity depends not only on the strength of inputs
but also on their number, this may induce a correlation between synaptic
strengths and in-degree. In line with this hypothesis, excitatory
postsynaptic currents (EPSCs) at single synapses were found to be
inversely related to the density of active synapses onto cultured
hippocampal neurons \cite{Liu95_404}, and the size of both miniature
EPSCs and evoked EPSCs between neurons decreased with network size
and with the number of synapses per neuron in patterned cultures \cite{Wilson07_13581},
although contrasting results have also been reported \cite{Ivenshitz10_1052,Medalla15_112}.
In the development of a mammal, the neuronal network grows by orders
of magnitude and is continuously modified. For instance, the amplitude
of miniature EPSCs is reduced in a period of heightened synaptogenesis
in rat primary visual cortex \cite{Desai2002_783}. During such developmental
processes, some functions are conserved and new functions emerge.
This balance between stability and flexibility is an intriguing theoretical
problem. Here, network scaling is deeply related to biological principles.
Our results open up a new perspective for analyzing and interpreting
such biological scaling laws. 

Certainly, most network models will not fit neatly into the categories
considered here, and detailed models often provide valuable insights
regardless of whether they are scaled in a systematic manner. Nevertheless,
it is usually possible to at least mention whether and how a particular
model is scaled. When the results are not amenable to mathematical
analysis, we suggest investigating through simulations of networks
of different sizes how essential characteristics depend on numbers
of neurons and synapses (the relevant characteristics depend on the
model at hand, and do not necessarily include mean activities or correlations).
Thus, while both the investigation of the infinity limit and the exploration
of downscaled networks remain powerful methods of computational neuroscience,
we argue for a more careful approach to network scaling than has hitherto
been customary, making the type of scaling and its consequences explicit.
Fortunately, in neuroscience full-scale simulations are now becoming
routinely possible due to the technological advances of recent years.

\section{Methods}

\paragraph*{Software}

We verify analytical results for networks of binary neurons and networks
of spiking neurons using direct simulations performed with NEST \cite{Gewaltig_07_11204}
revisions 10711 and 11264 for the spiking networks and revision 11540
for the binary networks. For simulating the multi-layer microcircuit
model, PyNN version 0.7.6 (revision 1312) \cite{Davison08_11} was
used with NEST 2.6.0 as back end, single-threaded on 12 MPI processes
on a high-performance cluster. All simulations have a time step of
$0.1\ms$. Spike times in the microcircuit model are constrained to
the grid. The other spiking network simulations use precise spike
timing \cite{Hanuschkin10_113}. In part, Sage was used for symbolic
linear algebra \cite{Sage}. Pre- and post-processing and numerical
analysis were performed with Python.

\subsection{Network structure and notation\label{sub:Network-structure-and-notation}}

For both the binary and the spiking networks, we derive analytical
results where both the number of populations $N_{\mathrm{pop}}$ and
the population-level connectivity are arbitrary. Specific examples
are given of networks with two populations (one excitatory, one inhibitory)
and with either population-specific or population-independent connectivities.
In addition, we discuss a multi-layer spiking cortical microcircuit
model consisting of $77,169$ neurons with approximately $3\times10^{8}$
synapses, with eight populations (2/3E, 2/3I, 4E, 4I, 5E, 5I, 6E,
6I) and population-specific connection probabilities \cite{Potjans14_785},
slightly adjusted to enhance the asynchrony of the activity. The adjustments
consist of replacing normally by lognormally distributed weights with
the same mean and with coefficient of variation $3$; and using $4.5$
instead of $4$ as the relative strength of synapses from 4I to 4E
compared to excitatory synaptic strengths. Besides distributed synaptic
strengths, the model has binomially distributed in- and out-degrees,
and normally distributed delays (clipped at the simulation time step),
thereby deviating from the assumptions of our analytic theory. It
thus serves to evaluate the robustness of our analytical results to
such deviations from the underlying assumptions. 

In all cases, pairs of populations are randomly connected. In the
binary and one- and two-population LIF network simulations, in-degrees
are fixed and multiple directed connections between pairs of neurons
(multapses) are disallowed. In the multi-layer microcircuit model,
in-degrees are distributed and multapses are allowed. In case of population-specific
connectivities, we denote the (unique or mean) in-degree for connections
from population $\beta$ to population $\alpha$ by $K_{\alpha\beta}$,
and synaptic strengths by $J_{\alpha\beta}$. Population sizes are
denoted by $N_{\alpha}$. For the example networks with population-independent
connection probability, we denote the size of the excitatory population
by $N$, the in-degree from excitatory neurons by $K=pN$, and the
size of the inhibitory relative to the excitatory population by $\gamma$,
so that the inhibitory in-degree is $\gamma K$. Synaptic strengths
are also taken to only depend on the source population, and are written
as $J$ for excitatory and $-gJ$ for inhibitory synapses.

\subsection{Binary network dynamics\label{sub:Binary-network-dynamics}}

We denote the activity of neuron $j$ by $n_{j}(t)$. The state $n_{j}(t)$
of a binary neuron is either $0$ or $1$, where $1$ indicates activity,
$0$ inactivity \cite{Ginzburg94,Buice09_377,Renart10_587}. The state
of the network of $N$ such neurons is described by a binary vector
$\mathbf{n}=(n_{1},\ldots,n_{N})\in\{0,1\}^{N}$. We denote the mean
activity by $\langle n_{j}(t)\rangle_{t}$, where the average $\langle\rangle_{t}$
is over time and realizations of the stochastic activity. The neuron
model shows stochastic transitions (at random points in time) between
the two states $0$ and $1$. In each infinitesimal interval $[t,t+\delta t)$,
each neuron in the network has the probability $\frac{1}{\tau}\delta t$
to be chosen for update \cite{Hopfield82}, where $\tau$ is the time
constant of the neuronal dynamics. We use an equivalent implementation
in which the time points of update are drawn independently for all
neurons. For a particular neuron, the sequence of update points has
exponentially distributed intervals with mean duration $\tau$, i.e.,
update times form a Poisson process with rate $\tau^{-1}$. The stochastic
update constitutes a source of noise in the system. Given that the
$j$-th neuron is selected for update, the probability to end in the
up-state ($n_{j}=1$) is determined by the gain function $F_{j}(\mathbf{n}(t))=\Theta(\sum_{k}J_{jk}n_{k}(t)-\theta)$
which in general depends on the activity $\mathbf{n}$ of all other
neurons. Here $\theta$ denotes the threshold of the neuron and $\Theta(x)$
the Heaviside function. The probability of ending in the down state
($n_{j}=0$) is $1-F_{j}(\mathbf{n})$. This model has been considered
previously \cite{Ginzburg94,Buice09_377,Hertz91}, and here we follow
the notation introduced in \cite{Buice09_377} that we also employed
in our earlier works. We skip details of the derivation here that
are already contained in \cite{Helias14}.

\subsection{First and second moments of activity in the binary network\label{sub:First-and-second-moments-binary}}

The combined distribution of large numbers of independent inputs can
be approximated as a Gaussian $\mathcal{N}(\mu,\sigma^{2})$ by the
central limit theorem. The arguments $\mu$ and $\sigma$ are the
mean and standard deviation of the synaptic input noise, together
referred to as the working point {[}cf. \eqref{eq:mu} and \eqref{eq:sigma}{]}.
The stationary mean activity of a given population of neurons then
obeys \cite{VanVreeswijk98_1321,Renart10_587,Grytskyy13_258,Helias14}
\begin{eqnarray}
\langle n\rangle & = & \langle F(\mathbf{n})\rangle\nonumber \\
 & \simeq & \int_{-\infty}^{\infty}\Theta(x-\theta)\mathcal{N}(\mu,\sigma^{2},x)dx\nonumber \\
 & = & \int_{\theta}^{\infty}\mathcal{N}(\mu,\sigma^{2},x)dx\nonumber \\
 & = & \frac{1}{2}\mathrm{erfc}\left(\frac{\theta-\mu\left(\langle\mathbf{n}\rangle\right)}{\sqrt{2}\sigma\left(\langle\mathbf{n}\rangle\right)}\right).\label{eq:binary_mean_activity}
\end{eqnarray}
This equation needs to be solved self-consistently because $\langle n\rangle$
influences $\mu,\sigma$ through interactions within the population
itself and with other populations.

When network activity is stationary, the covariance of the activities
of a pair $(j,k)$ of neurons is defined as $c_{jk}(\Delta)=\langle\delta n_{j}(t+\Delta)\delta n_{k}(t)\rangle_{t}$,
where $\delta n_{j}(t)=n_{j}(t)-\langle n_{j}(t)\rangle_{t}$ is the
deviation of neuron $j$'s activity from expectation, and $\Delta$
is a time lag. Instead of the raw correlation $\langle n_{j}(t+\Delta)n_{k}(t)\rangle_{t}$,
here and for the spiking networks we measure the covariance, i.e.,
the second centralized moment, which is also identical to the second
cumulant. To derive analytical expressions for the covariances in
binary networks in the asynchronous regime, we follow the theory developed
in \cite{Ginzburg94,Renart10_587,Grytskyy13_258,Grytskyy13_131,Helias14}.
We first consider the case of vanishing transmission delays $d=0$
and then discuss networks with delays. 

Let 
\begin{equation}
c_{\alpha\beta}=\frac{1}{N_{\alpha}N_{\beta}}\;\sum_{j\in\alpha,k\in\beta,j\neq k}c_{jk}\label{eq:c_alpha_beta}
\end{equation}
be the covariance averaged over disjoint pairs of neurons in two (possibly
identical) populations $\alpha,\beta$, and $a_{\alpha}=\frac{1}{N_{\alpha}}\sum_{j\in\alpha}a_{j}$
the population-averaged single-neuron variance $a_{j}(\Delta)=\langle\delta n_{j}(t+\Delta)\delta n_{j}(t)\rangle_{t}$.
Note that for $\alpha=\beta$ there are only $N_{\alpha}(N_{\alpha}-1)$
disjoint pairs of neurons, so\textrm{ $c_{\alpha\alpha}$} differs
from the average pairwise cross-correlation by a factor $(N_{\alpha}-1)/N_{\alpha}$,
but we choose this definition because it slightly simplifies the population-level
equations. For sufficiently weak synapses and sufficiently high firing
rates, and when higher-order correlations can be neglected, a linearized
equation relating these quantities can be derived for the case $d=0$
(\cite{Ginzburg94} eqs. (9.14)--(9.16); \cite{Renart10_587} supplementary
material eq. (36), \cite{Helias14} eq. (10)),
\begin{equation}
2c_{\alpha\beta}=\sum_{\gamma}\left(W_{\alpha\gamma}c_{\gamma\beta}+W_{\beta\gamma}c_{\gamma\alpha}\right)+W_{\alpha\beta}\frac{a_{\beta}}{N_{\beta}}+W_{\beta\alpha}\frac{a_{\alpha}}{N_{\alpha}}.\label{eq:corr_self_consistent_pop}
\end{equation}
Here, we have assumed identical time constants across populations,
and

\begin{equation}
W_{\alpha\beta}=S(\mu_{\alpha},\sigma_{\alpha})J_{\alpha\beta}K_{\alpha\beta}\label{eq:binary_effective_connectivity}
\end{equation}
is the linearized effective connectivity. The susceptibility $S$
is defined as the slope of the gain function averaged over the noisy
input to each neuron \cite{Grytskyy13_258,Grytskyy13_131,Helias14},
reducing for a Heaviside gain function to
\begin{equation}
S(\mu,\sigma)=\frac{1}{\sqrt{2\pi}\sigma}e^{-\frac{(\mu-\theta)^{2}}{2\sigma^{2}}}.\label{eq:susceptibility}
\end{equation}

With the definitions
\begin{equation}
\bar{c}_{\alpha\beta}\equiv\frac{1}{N_{\alpha}N_{\beta}}\;\sum_{j\in\alpha,k\in\beta}c_{jk}=c_{\alpha\beta}+\delta_{\alpha\beta}\frac{a_{\alpha}}{N_{\alpha}}\label{eq:c_cbar}
\end{equation}
\begin{equation}
P_{\alpha\beta}\equiv\delta_{\alpha\beta}-W_{\alpha\beta}
\end{equation}
 \eqref{eq:corr_self_consistent_pop} is recognized as a continuous
Lyapunov equation
\begin{equation}
\mathbf{P\bar{\mathbf{c}}}+(\mathbf{P\bar{\mathbf{c}}})^{T}=2\mathrm{diag}\left(\frac{a_{\alpha}}{N_{\alpha}}\right)\equiv2\mathbf{A},\label{eq:Lyapunov}
\end{equation}
which can be solved using known methods. Let $\mathbf{v}^{j},\mathbf{u}^{k}$
be the left and right eigenvectors of $\mathbf{W}$, with eigenvalues
$\lambda_{j}$ and $\lambda_{k}$, respectively. Choose the normalization
such that the left and right eigenvectors are biorthogonal, 
\begin{equation}
\mathbf{v}^{jT}\mathbf{u}^{k}=\delta_{jk}.\label{eq:biorthogonality}
\end{equation}
Then multiplying \eqref{eq:Lyapunov} from the left with $\mathbf{v}^{jT}$
and from the right with $\mathbf{v}^{k}$ yields
\begin{equation}
(1-\lambda_{j})\mathbf{v}^{jT}\mathbf{\bar{c}}\mathbf{v}^{k}+\mathbf{v}^{jT}\mathbf{\bar{c}}\mathbf{v}^{k}(1-\lambda_{k})=2\mathbf{v}^{jT}\mathbf{A}\mathbf{v}^{k}.\label{eq:Lyapunov_projected}
\end{equation}
Define

\begin{equation}
m^{jk}\equiv\mathbf{v}^{jT}\mathbf{m}\mathbf{v}^{k},\label{eq:c_ik}
\end{equation}
for $\mathbf{m}=\mathbf{c},\bar{\mathbf{c}},\mathbf{A}$. Then solving
\eqref{eq:Lyapunov_projected} for $\mathbf{\bar{c}}$ gives
\begin{equation}
\mathbf{\bar{c}=}\sum_{j,k}\frac{2A^{jk}}{2-\lambda_{j}-\lambda_{k}}\mathbf{u}^{j}\mathbf{u}^{kT},\label{eq:c_bar1}
\end{equation}
as can be verified using \eqref{eq:biorthogonality}. This provides
an approximation of the population-averaged zero-lag correlations,
including contributions from both auto- and cross-correlations.

To determine the temporal structure of the population-averaged cross-correlations,
we start from the single-neuron level, for which the correlations
approximately obey (\cite{Grytskyy13_131} eq. (29))

\begin{equation}
\tau\frac{d}{d\Delta}c_{jk}(\Delta)+c_{jk}(\Delta)=\sum_{i}w_{ji}c_{ik}(\Delta),\qquad\Delta\geq0,\label{eq:binary_corr_evolution}
\end{equation}
where $w_{ij}$ is the neuron-level effective connectivity ($w_{ij}=S_{i}J_{ij}$
if a connection exists and $w_{ij}=0$ otherwise). This equation also
holds on the diagonal, $j=k$. To obtain the population-level equation,
we use \eqref{eq:c_alpha_beta} and \eqref{eq:c_cbar} and count the
numbers of connections, which yields a factor $K_{\alpha\beta}$ 
 for each projection. Equation \eqref{eq:binary_corr_evolution} then
becomes
\begin{equation}
\tau\frac{d}{d\Delta}\mathbf{\bar{\mathbf{c}}}(\Delta)=-\mathbf{P}\mathbf{\bar{\mathbf{c}}}(\Delta),\qquad\Delta\geq0.\label{eq:corr_differential}
\end{equation}
This step from the single-neuron to the population level constitutes
an approximation when the out-degrees are distributed, but is exact
for fixed out-degree \cite{Tetzlaff12_e1002596,Grytskyy13_131}. The
correlations for $\Delta<0$ are determined by $\bar{c}_{\alpha\beta}(-\Delta)=\bar{c}_{\beta\alpha}(\Delta)$.
With the definition \eqref{eq:c_ik}, \eqref{eq:corr_differential}
yields 
\begin{equation}
\tau\frac{d}{d\Delta}\bar{c}^{jk}(\Delta)=(\lambda_{j}-1)\bar{c}^{jk}(\Delta)\qquad\Delta\geq0.\label{eq:binary_pop_corr}
\end{equation}
Using the initial condition for $\bar{\mathbf{c}}$ from \eqref{eq:c_bar1}
and multiplying \eqref{eq:binary_pop_corr} by $\mathbf{u}^{j}\mathbf{u}^{kT}$,
summing over $j$ and $k$, we obtain the solution

\begin{equation}
\mathbf{\bar{\mathbf{c}}}(\Delta\geq0)=\sum_{j,k}\frac{2A^{jk}}{2-\lambda_{j}-\lambda_{k}}\mathbf{u}^{j}\mathbf{u}^{kT}e^{\frac{\lambda_{j}-1}{\tau}\Delta}.\label{eq:binary_cov_matrix_vs_time-1}
\end{equation}
The shape of the autocovariances is well approximated by that for
isolated neurons, $\mathbf{A}e^{-\frac{\Delta}{\tau}}$, with corrections
due to interactions being $O(1/N)$ \cite{Ginzburg94}. Substituting
this form in \eqref{eq:binary_cov_matrix_vs_time-1} leads to

\begin{equation}
\mathbf{c}(\Delta\geq0)=\sum_{j,k}\frac{2A^{jk}}{2-\lambda_{j}-\lambda_{k}}\mathbf{u}^{j}\mathbf{u}^{kT}e^{\frac{\lambda_{j}-1}{\tau}\Delta}-\mathbf{A}e^{-\frac{\Delta}{\tau}},\label{eq:binary_cov_matrix_vs_time}
\end{equation}
equivalent to \cite{Ginzburg94} eq. (6.20). Note that this equation
still needs to be solved self-consistently, because the variance of
the inputs to the neurons, which goes into $S(\mu,\sigma)$, depends
on the correlations. However, correlations tend to contribute only
a small fraction of the input variance in the asynchronous regime
(cf. \cite{Helias14} Fig. 3D). The accuracy of the result \eqref{eq:binary_cov_matrix_vs_time}
is illustrated in \prettyref{fig:binary_scaling_asymm}A for a network
with parameters given in \prettyref{tab:binary_network-1} by comparison
with a direct simulation. Note that the delays were not zero but equal
to the simulation time step of $0.1\ms$, sufficiently small for the
correlations to be well approximated by \eqref{eq:binary_cov_matrix_vs_time}.

Now consider arbitrary transmission delay $d>0$, and let both $d$
and the input statistics be population-independent. This case is most
easily approached from the Fourier domain, where the population-averaged
covariances including autocovariances can be approximated as \cite{Grytskyy13_131}
\begin{equation}
\mathbf{\bar{\mathbf{C}}}(\omega)=\left(H(\omega)^{-1}-\mathbf{W}\right)^{-1}2\tau\mathbf{A}\left(H(-\omega)^{-1}-\mathbf{W}^{T}\right)^{-1}.\label{eq:binary_corr_fourier}
\end{equation}
Here, $H(\omega)$ is the transfer function
\begin{equation}
H(\omega)=\frac{e^{-i\omega d}}{1+i\omega\tau},\label{eq:transfer_fn}
\end{equation}
which is equal for all populations under the assumptions made. The
transfer function is the Fourier transform of the impulse response,
which is a jump followed by an exponential relaxation, 
\begin{equation}
h(t)=\Theta(t-d)\frac{1}{\tau}e^{-\frac{t-d}{\tau}},\label{eq:exp_kernel}
\end{equation}
where $\Theta$ is the Heaviside step function.

For the case of population-independent $H(\omega)$, Fourier back
transformation to the time domain is feasible, and was performed in
\cite{Grytskyy13_131} for symmetric connectivity matrices. Here,
we consider generic connectivity (insofar as consistent with equal
$H(\omega)$), and again use projection onto the eigenspaces of $\mathbf{W}$\textrm{
}to obtain a form similar to \eqref{eq:binary_cov_matrix_vs_time},
i.e., insert the identity matrix\textrm{
\begin{equation}
\sum_{j}\mathbf{u}^{j}\mathbf{v}^{jT}=\unit{\mathbf{1}\!\!\mathbf{1}}
\end{equation}
}both on the left and on the right of \eqref{eq:binary_corr_fourier},
and Fourier transform to obtain
\begin{align}
2\pi\mathbf{\bar{\mathbf{c}}}(\Delta) & =\intop_{-\infty}^{+\infty}\mathbf{\bar{\mathbf{C}}}(\omega)e^{i\omega\Delta}d\omega\nonumber \\
 & =\intop_{-\infty}^{+\infty}e^{i\omega\Delta}\sum_{j,k}\mathbf{u}^{j}\frac{1}{H(\omega)^{-1}-\lambda_{j}}2\tau\mathbf{v}^{jT}\mathbf{A}\mathbf{v}^{k}\frac{1}{H(-\omega)^{-1}-\lambda_{k}}\mathbf{u}^{kT}d\omega\nonumber \\
 & =2\tau\sum_{j,k}\mathbf{u}^{j}\mathbf{u}^{kT}\, A^{jk}\,\intop_{-\infty}^{+\infty}f_{jk}(\omega)\, e^{i\omega\Delta}\, d\omega\nonumber \\
\mathrm{with}\; f_{jk}(\omega) & \equiv\frac{1}{H(\omega)^{-1}-\lambda_{j}}\,\frac{1}{H(-\omega)^{-1}-\lambda_{k}}.\label{eq:binary_corr_transform}
\end{align}
In the third line of \eqref{eq:binary_corr_transform}, we used $A^{jk}=\mathbf{v}^{jT}\mathbf{A}\mathbf{v}^{k}$
and collected the frequency-dependent terms for clarity. The exponential
$e^{i\omega\Delta}$ does not have any poles, so the only poles stem
from $f_{jk}$, which we denote by $z_{l}\left(\lambda_{j}\right)$
and the corresponding residues by $\mathrm{Res}_{j,k}\left[z_{l}\left(\lambda_{j}\right)\right]$.
We only need to consider $\Delta\geq0$, since the solution for negative
lags follows from $\bar{\mathbf{c}}(\Delta)=\bar{\mathbf{c}}^{T}(-\Delta)$.
The equation can then be solved by contour integration over the upper
half of the complex plane, as the integrand vanishes at $\omega\to+i\infty$.
Stability requires that the poles of the first term of \eqref{eq:binary_corr_transform}
lie only in the upper half plane (note that the linear approximation
we have employed only applies in the stable regime). The poles of
the second term correspondingly lie in the lower half plane and hence
need not be considered. For $d>0$, the locations of the poles are
given by \cite{Grytskyy13_131} eq. (12),
\begin{equation}
z_{l}(\lambda_{j})=\frac{i}{\tau}-\frac{i}{d}W_{l}\left(\lambda_{j}\frac{d}{\tau}e^{d/\tau}\right),\label{eq:Lambert}
\end{equation}
where $W_{l}$ is the $l^{\mathrm{th}}$ of the infinitely many branches
of the Lambert-W function defined by $x=W(x)e^{W(x)}$ \cite{Corless96_329}.
For $d=0$, the poles are $z(\lambda_{j})=-\frac{i}{\tau}\left(\lambda_{j}-1\right)$.
Using the residue theorem thus brings \eqref{eq:binary_corr_transform}
into the form

\begin{eqnarray}
\mathbf{\bar{\mathbf{c}}}(\Delta\geq0) & = & 2\tau iI(\gamma)\sum_{j,k,l}\mathbf{u}^{j}\mathbf{u}^{kT}\, A^{jk}\mathrm{Res}_{j,k}\left[z_{l}\left(\lambda_{j}\right)\right]e^{iz_{l}\left(\lambda_{j}\right)\Delta}\nonumber \\
 & = & \sum_{j,k,l}a_{jkl}\mathbf{u}^{j}\mathbf{u}^{kT}e^{iz_{l}\left(\lambda_{j}\right)\Delta},\nonumber \\
\mathrm{with\;}a_{jkl} & \equiv & 2\tau iI(\gamma)A^{jk}\mathrm{Res}_{j,k}\left[z_{l}\left(\lambda_{j}\right)\right],\label{eq:residue_theorem}
\end{eqnarray}
where\textrm{ }$I(\gamma)=1$ is the winding number of the contour
$\gamma$ around the poles. To see that \eqref{eq:residue_theorem}
reduces to \eqref{eq:binary_cov_matrix_vs_time} when $d=0$, substitute
the poles in the upper half plane $z(\lambda_{j})=-\frac{i}{\tau}\left(\lambda_{j}-1\right)$
with residues $\left[i\tau\left(2-\lambda_{j}-\lambda_{k}\right)\right]^{-1}$
and note that $\mathbf{c}(\Delta)=\bar{\mathbf{c}}(\Delta)-\mathbf{A}(\Delta)$. 

When the input statistics and hence transfer functions are population-specific,
\eqref{eq:binary_corr_fourier} becomes
\begin{eqnarray}
\bar{\mathbf{C}}(\omega) & = & \left(\mathbf{1}\!\!\mathbf{1}-\mathbf{M}(\omega)\right)^{-1}\mathbf{D}(\omega)\left(\mathbf{1}\!\!\mathbf{1}-\mathbf{M}^{T}(-\omega)\right)^{-1},\label{eq:binary_cov_matrix_general_H}\\
\mathbf{D}(\omega) & \equiv & \mathrm{diag}\left(\left\{ \frac{2\tau_{\alpha}a_{\alpha}}{N_{\alpha}\left(1+\omega^{2}\tau_{\alpha}^{2}\right)}\right\} _{\alpha=1\ldots N_{\mathrm{pop}}}\right),
\end{eqnarray}
where \textrm{$M_{\alpha\beta}(\omega)=H_{\alpha\beta}(\omega)W_{\alpha\beta}$.}

\subsection{Spiking network dynamics\label{sub:Spiking-network-dynamics}}

The spiking networks consist of single-compartment leaky integrate-and-fire
neurons with exponential current-based synapses. The subthreshold
dynamics of neuron $i$ is given by

\begin{eqnarray}
\tau_{\mathrm{m}}\frac{dV_{i}}{dt} & = & -V_{i}+I_{i}(t),\nonumber \\
\tau_{\mathrm{s}}\frac{dI_{i}}{dt} & = & -I_{i}+\tau_{\mathrm{m}}\sum_{j}J_{ij}s_{j}(t-d),\label{eq:LIF_dynamics-1}
\end{eqnarray}
where we have set the resting potential to zero without loss of generality,
and absorbed the membrane resistance into the synaptic current $I_{i}$,
in line with previous works \cite{Fourcaud02,Helias13_023002}. Bringing
back the corresponding parameters, the dynamics reads

\begin{eqnarray}
\tau_{\mathrm{m}}\frac{d\tilde{V}_{i}}{dt} & = & -\left(\tilde{V}_{i}-E_{\mathrm{L}}\right)+R_{\mathrm{m}}\tilde{I}_{i}(t),\nonumber \\
\tau_{\mathrm{s}}\frac{d\tilde{I}_{i}}{dt} & = & -\tilde{I}_{i}+\tau_{\mathrm{s}}\sum_{j}\tilde{J}_{ij}s_{j}(t-d).\label{eq:LIF_dynamics-2}
\end{eqnarray}
Thus, our scaled synaptic amplitudes $J_{ij}$ in terms of the amplitudes
$\tilde{J}_{ij}$ of the synaptic current due to a single spike are
$J_{ij}=R_{\mathrm{m}}\tau_{\mathrm{s}}/\tau_{\mathrm{m}}\tilde{J}_{ij}$.
Here, $\tau_{\mathrm{m}}$ and $\tau_{\mathrm{s}}$ are membrane and
synaptic time constants, $E_{\mathrm{L}}$ is the leak or resting
potential, $R_{\mathrm{m}}$ is the membrane resistance, $d$ is the
transmission delay, $\tilde{I}_{i}=I_{i}/R_{\mathrm{m}}$ is the total
synaptic current, and $s_{j}=\sum_{k}\delta(t-t_{k}^{j})$ are the
incoming spike trains. When $V_{i}$ reaches a threshold $\theta$,
a spike is assumed, and the membrane potential is clamped to a level
$V_{\mathrm{r}}$ for a refractory period $\tau_{\mathrm{ref}}$.
Threshold and reset potential in physical units are shifted by the
leak potential $\theta=\tilde{\theta}-E_{\mathrm{L}}$, $V_{\mathrm{r}}=\tilde{V_{\mathrm{r}}}-E_{\mathrm{L}}$,
showing that the assumption $E_{\mathrm{L}}=0$ in \eqref{eq:LIF_dynamics-1}
does not limit generality. The intrinsic dynamics of the neurons in
the different populations are taken to be identical, so that population
differences are only expressed in the couplings.

\subsection{First and second moments of activity in the spiking network\label{sub:LIF_moments}}

An approximation of the stationary mean firing rate of LIF networks
with exponential current-based synapses was derived in \cite{Fourcaud02},
\begin{eqnarray}
r & = & \left(\tau_{\mathrm{m}}\sqrt{\pi}\int_{\frac{V_{r}-\mu}{\sigma}+\frac{\alpha}{2}\sqrt{\frac{\tau_{\mathrm{s}}}{\tau_{\mathrm{m}}}}}^{\frac{\theta-\mu}{\sigma}+\frac{\alpha}{2}\sqrt{\frac{\tau_{\mathrm{s}}}{\tau_{\mathrm{m}}}}}\Psi(s)\, ds\right)^{-1},\nonumber \\
\Psi(s) & = & e^{s^{2}}\left(1+\mathrm{erf}(s)\right),\nonumber \\
\alpha & = & \sqrt{2}\left|\zeta\left(\frac{1}{2}\right)\right|,\label{eq:siegert}
\end{eqnarray}
where the summed synaptic input is characterized by a Gaussian noise
with first moment $\mu$ and second moment $\sigma^{2}$ based on
the diffusion approximation, and $\zeta$ is the Riemann zeta function.

For the covariances, we follow and extend the theory developed in
\cite{Helias13_023002,Grytskyy13_131}, starting with the average
influence of a single synapse. Assuming that the network is in the
asynchronous state, and that synaptic amplitudes are small, the synaptic
influences can be averaged around the mean activity $r_{j}$ of each
neuron $j$. These influences are characterized by linear response
kernels $h_{jk}(t,t^{\prime})$ defined as the derivative of the density
of spikes of spike train $s_{j}(t)$ of neuron $j$ with respect to
an incoming spike train $s_{k}(t^{\prime})$, averaged over realizations
of the remaining incoming spike trains $\mathbf{s\backslash}s_{k}$
that act as noise. In the stationary state, the kernel only depends
on the time difference $t-t^{\prime}$, giving
\[
\langle s_{j}(t)|s_{k}\rangle_{\mathbf{s}\backslash s_{k}}=r_{j}+\int_{-\infty}^{t}h_{jk}(t-t^{\prime})(s_{k}(t^{\prime})-r_{k})\, dt^{\prime},
\]
\begin{eqnarray}
h_{jk}(t-t^{\prime}) & = & \left\langle \frac{\delta s_{j}(t)}{\delta s_{k}(t^{\prime})}\right\rangle _{\mathbf{s}\backslash s_{k}}\nonumber \\
 & \equiv & w_{jk}h(t-t^{\prime}),\label{eq:LIF_kernel}
\end{eqnarray}
where $\delta s_{j}\equiv s_{j}-r_{j}$ is the $j$-th centralized
(zero mean) spike train. Here, $w_{jk}$ is the integral of $h_{jk}(t-t^{\prime})$,
and $h(t-t^{\prime})$ is a normalized function capturing its time
dependence, which may be source- and target-specific. The dimensionless
effective weights $w_{jk}$ are determined nonlinearly by the synaptic
strengths $J_{jk}$, the single-neuron parameters, and the working
point $(\mu_{j},\sigma_{j})$ (cf. \cite{Helias13_023002} eq. (A.3)
but note that $\beta$ as given there has a spurious factor $J$).
We approximate the impulse response by the form \eqref{eq:exp_kernel},
where $\tau$ is now an effective time constant depending on the working
point $(\mu_{j},\sigma_{j})$ and the parameters of the target neurons.
This form of the impulse response, corresponding to a low-pass filter,
appears to be a good approximation in the noisy regime when the neuron
fires irregularly. In the mean-driven regime ($\mu\gg\sigma$) the
transfer function of the LIF neuron is known to exhibit resonant behavior
with a peak close to its firing rate. In this regime a single exponential
response kernel is expected to be a poor approximation (see, e.g.,
\cite{Brunel01_2186} Fig. 1). In general, the source population dependence
of \eqref{eq:exp_kernel} comes in through the delay $d$, and the
target population dependence through both $\tau$ and $d$. 

As for binary networks with delays, the average pairwise covariance
functions $c_{ij}(\Delta)\equiv\langle\delta s_{i}(t+\Delta)\delta s_{j}(t)\rangle_{t}$
are most conveniently derived starting from the frequency domain.
In case of identical transfer functions for all populations, the matrix
of average cross-covariances is given by \cite{Grytskyy13_131} eq.
(16) minus the autocovariance contribution,

\begin{eqnarray}
\mathbf{C}(\omega) & = & \left(H(\omega)^{-1}-\mathbf{W}\right)^{-1}\mathbf{W}\mathbf{A}\mathbf{W}^{T}\left(H(-\omega)^{-1}-\mathbf{W}^{T}\right)^{-1}\nonumber \\
 & + & \left(H(\omega)^{-1}-\mathbf{W}\right)^{-1}\mathbf{WA}\nonumber \\
 & + & \mathbf{A}\mathbf{W}^{T}\left(H(-\omega)^{-1}-\mathbf{W}^{T}\right)^{-1}.\label{eq:LIF_corr_symm}
\end{eqnarray}
\begin{comment}
\begin{equation}
\mathbf{C}(\omega)\!=\!\mathbf{P}(\omega)\!\left(\mathbf{M}(\omega)\mathbf{A}\!+\mathbf{A}\!\mathbf{M}^{T}(-\omega)\!-\!\mathbf{M}(\omega)\mathbf{A}\mathbf{M}^{T}(-\omega)\right)\!\mathbf{P}^{T}(-\omega),\label{eq:corr_generic_spiking-1}
\end{equation}
\end{comment}
Here, $\mathbf{\mathbf{W}}$ contains the effective weights of single
synapses from population $\beta$ to population $\alpha$ times the
corresponding in-degrees, $w_{\alpha\beta}K_{\alpha\beta}$; and $\mathbf{A}$
contains the population-averaged autocovariances, which we approximate
as \textrm{$\delta_{\alpha\beta}\frac{r_{\alpha}}{N_{\alpha}}$},
with $r_{\alpha}$ the mean firing rate, as also done in \cite{Helias13_023002}.
In \cite{Grytskyy13_131}, \eqref{eq:LIF_corr_symm} was written using
a more general diagonal matrix instead of $\mathbf{A}$, to help clarify
close similarities between binary and LIF networks and Ornstein-Uhlenbeck
processes or linear rate models; however, for LIF networks, this diagonal
matrix corresponds precisely to the autocovariance matrix. We chose
the form \eqref{eq:LIF_corr_symm} because it separates terms that
vanish at either $\omega\to i\infty$ or $\omega\to-i\infty$ depending
on $\Delta$. This facilitates Fourier back transformation, as contour
integration with an appropriate contour can be used for each term.

To perform the Fourier back transformation, we apply the same method
as used for the binary network. Let \textrm{$\mathbf{v}^{j},\mathbf{u}^{j}$
}be the left and right eigenvectors of the connectivity matrix $\mathbf{W}$,
and $\lambda_{j}$ the corresponding eigenvalues. Insert $\sum_{j}\mathbf{u}^{j}\mathbf{v}^{jT}=\unit{\mathbf{1}\!\!\mathbf{1}}$
into \eqref{eq:LIF_corr_symm} on the left and right, and Fourier
transform,

\begin{align}
2\pi\mathbf{c}(\Delta) & =\intop_{-\infty}^{+\infty}\mathbf{C}(\omega)e^{i\omega\Delta}d\omega\nonumber \\
 & =\intop_{-\infty}^{+\infty}e^{i\omega\Delta}\Bigg\{\sum_{j,k}\mathbf{u}^{j}\frac{\lambda_{j}}{H(\omega)^{-1}-\lambda_{j}}\mathbf{v}^{jT}\mathbf{A}\mathbf{v}^{k}\frac{\lambda_{k}}{H(-\omega)^{-1}-\lambda_{k}}\mathbf{u}^{kT}\nonumber \\
 & +\sum_{j,k}\mathbf{u}^{j}\frac{\lambda_{j}}{H(\omega)^{-1}-\lambda_{j}}\mathbf{v}^{jT}\mathbf{A}\mathbf{v}^{k}\mathbf{u}^{kT}\nonumber \\
 & +\sum_{j,k}\mathbf{u}^{j}\mathbf{v}^{jT}\mathbf{A}\mathbf{v}^{k}\frac{\lambda_{k}}{H(-\omega)^{-1}-\lambda_{k}}\mathbf{u}^{kT}\Bigg\} d\omega.\label{eq:LIF_corr_Fourier_transform}
\end{align}
As for the binary case, we only need to consider $\Delta\geq0$, as
the solution for $\Delta<0$ is given by $\mathbf{c}(\Delta)=\mathbf{c}^{T}(-\Delta)$.
The contour can then be closed over the upper half plane, where the
term containing only $H(-\omega)$ has no poles due to the stability
condition. When $\Delta<d$, the contour for the term containing only
$H(\omega)$ can also be closed in the lower half plane where it
has no poles, so that the corresponding integral vanishes. Analogously,
the integral of the term with only $H(-\omega)$ vanishes when $0>\Delta>-d$.
Therefore, the second and third terms represent `echoes' of spikes
arriving after one transmission delay \cite{Grytskyy13_131}. For
$\Delta=0$ and $d>0$, only the first term contributes, and the contour
can be closed in either half plane. As before, the poles are given
by \eqref{eq:Lambert} for $d>0$, and by $z(\lambda_{j})=\mp\frac{i}{\tau}\left(\lambda_{j}-1\right)$
for $d=0$. The residue theorem yields a solution of the form \eqref{eq:residue_theorem},
the only difference being the precise form of the residues, and the
fact that we here consider $\mathbf{c}$ as opposed to $\bar{\mathbf{c}}$. 

In the absence of delays, an explicit solution can again be derived.
For $\Delta>0$, the poles inside the contour are $z(\lambda_{j})=-\frac{i}{\tau}(\lambda_{j}-1)$
corresponding to the terms with \textrm{$H(\omega)^{-1}$}. The residue
corresponding to $\frac{\lambda_{j}}{H(\omega)^{-1}-\lambda_{j}}$
is $\frac{\lambda_{j}}{i\tau}$, and the term $\frac{\lambda_{k}}{H(-\omega)^{-1}-\lambda_{k}}$
is finite and evaluates at the pole to $\frac{\lambda_{k}}{2-\lambda_{j}-\lambda_{k}}$.
Using $A^{jk}=\mathbf{v}^{jT}\mathbf{A}\mathbf{v}^{k}$ we get 
\begin{equation}
\mathbf{c}(\Delta>0)=\sum_{j,k}\frac{A^{jk}}{\tau}\frac{\lambda_{j}\left(2-\lambda_{j}\right)}{2-\lambda_{j}-\lambda_{k}}\mathbf{u}^{j}\mathbf{u}^{kT}e^{\frac{\lambda_{j}-1}{\tau}\Delta},\label{eq:LIF_cov_matrix}
\end{equation}
which is reminiscent of but not identical to \eqref{eq:binary_cov_matrix_vs_time}
for the binary network. Note that \eqref{eq:LIF_cov_matrix} for the
LIF network corresponds to spike train covariances with the dimensionality
of $1/t^{2}$ due to $[A^{jk}]=[1/t]$ and the factor $1/\tau$, whereas
the covariances for the binary network are dimensionless.

The population-specific generalization of \eqref{eq:LIF_corr_symm}
reads

\begin{eqnarray}
\mathbf{\mathbf{C}}(\omega) & = & \left(\mathbf{\mathbf{1}\!\!\mathbf{1}}-\mathbf{M}(\omega)\right)^{-1}\mathbf{M}(\omega)\mathbf{A\mathbf{M}}^{T}(-\omega)\left(\mathbf{\mathbf{1}\!\!\mathbf{1}}-\mathbf{M}^{T}(-\omega)\right)^{-1}\nonumber \\
 & + & \left(\mathbf{\mathbf{1}\!\!\mathbf{1}}-\mathbf{M}(\omega)\right)^{-1}\mathbf{M}(\omega)\mathbf{A}\nonumber \\
 & + & \mathbf{A}\mathbf{M}^{T}(-\omega)\left(\mathbf{\mathbf{1}\!\!\mathbf{1}}-\mathbf{M}^{T}(-\omega)\right)^{-1},
\end{eqnarray}
where $\mathbf{M}(\omega)$ has elements\textrm{ }$H_{\alpha\beta}(\omega)K_{\alpha\beta}w_{\alpha\beta}$,
as before. The covariance matrix including autocovariances can be
more simply written as 
\begin{equation}
\mathbf{\bar{\mathbf{C}}}(\omega)=\left(\mathbf{\mathbf{1}\!\!\mathbf{1}}-\mathbf{M}(\omega)\right)^{-1}\mathbf{A}\left(\mathbf{\mathbf{1}\!\!\mathbf{1}}-\mathbf{M}^{T}(-\omega)\right)^{-1}.\label{eq:LIF_corr_simple}
\end{equation}
The only difference compared to the expression \eqref{eq:binary_cov_matrix_general_H}
for the binary network is the form of the diagonal matrix, here analogous
to white output noise in a linear rate model, whereas the binary network
resembles a linear rate model with white noise on the input side,
which is passed through the transfer function before affecting the
correlations \cite{Grytskyy13_131}. %

\subsection{Fluctuating rate equation and stability condition\label{sub:Fluctuating-rate-equation-1}}

An alternative description of the spiking dynamics can be obtained
by considering a system of linear coupled rate equations that produces
the same moments to second order as the spiking dynamics \cite{Grytskyy13_131}.
The convolution equation
\begin{eqnarray}
y_{j}(t) & = & \sum_{k}\int\, h_{jk}(t-t^{\prime})y_{k}(t^{\prime})\, dt^{\prime}+x_{j}(t)\label{eq:rate_dynamics}\\
\text{with }\nonumber \\
\langle x_{j}(t)x_{k}(s)\rangle & = & \delta_{jk}\, r_{j}\,\delta(t-s),\nonumber 
\end{eqnarray}
with pairwise uncorrelated white noises $x_{j}$ and the response
kernel $h_{jk}$ given by \eqref{eq:LIF_kernel} can be shown to yield
a cross-covariance matrix of the form \eqref{eq:LIF_corr_symm} by
considering the Fourier transform of \eqref{eq:rate_dynamics}, written
in matrix notation as
\begin{eqnarray}
\mathbf{Y}(\omega) & = & H(\omega)\,\mathbf{W}\,\mathbf{Y}(\omega)+\mathbf{X}(\omega).\label{eq:rate_dynamics_Fourier}
\end{eqnarray}
We can expand the latter equation into eigenmodes by multiplying from
the left with the left-sided eigenvector $\mathbf{v}^{k}$ of $\mathbf{W}$
and by writing the general solution as a linear combination of right-sided
eigenmodes $\mathbf{Y}(\omega)=\sum_{j}\eta_{j}(\omega)\,\mathbf{u}^{j}$
to obtain (with the bi-orthogonality relation $\mathbf{v}^{kT}\mathbf{u}^{j}=\delta_{kj}$)

\begin{eqnarray}
\eta_{k}(\omega) & = & H(\omega)\,\lambda_{k}\eta_{k}(\omega)+\mathbf{v}^{kT}\mathbf{X}(\omega)\nonumber \\
\eta_{k}(\omega) & = & \frac{1}{1-\lambda_{k}H(\omega)}\,\mathbf{v}^{kT}\mathbf{X}(\omega).\label{eq:rate_equation_projected}
\end{eqnarray}
The latter equation shows that the same poles $z(\lambda_{k})$ that
appear in the covariance function \eqref{eq:LIF_corr_Fourier_transform}
also determine the evolution of the effective rate equation. Moreover,
transforming \eqref{eq:rate_equation_projected} back to the time
domain, we see with
\begin{eqnarray*}
\eta_{k}(t) & =i & \sum_{\text{poles }z(\lambda_{k})}\mathrm{Res}\left(\frac{1}{1-\lambda_{k}H(z)},\, z(\lambda_{k})\right)\,\mathbf{v}^{k}\mathbf{X}(z)\, e^{\, iz(\lambda_{k})\, t}
\end{eqnarray*}
that the eigenmodes have a time evolution determined by $e^{iz(\lambda_{k})t}$.
Hence the imaginary part of the pole $z(\lambda_{k})$ controls whether
the mode is exponentially growing $\left(\mathrm{Im}(z)<0\right)$
or decaying $\left(\mathrm{Im}(z)>0\right)$, while the real part
determines the oscillation frequency.

\section{Acknowledgements}

We acknowledge funding by the Helmholtz Association: portfolio theme
Supercomputing and Modeling for the Human Brain (SMHB) and Helmholtz
young investigator group VH-NG-1028; and European Union Grants
269921 (BrainScaleS) and 604102 (Human Brain Project, HBP).

\bibliographystyle{plos2009_brain}
%\phantomsection\addcontentsline{toc}{section}{\refname}\bibliography{brain,math,computer}

\newpage{}

\section{Supporting information\label{sec:supplement}}

\subsection{Representation of eigenvalues and corresponding time scales in the
correlations: $\mathbf{d=0}$\label{sub:Representation-of-eigenvalues}}

We here show for networks with an identical transfer function across
populations and without transmission delays that, apart from potential
vanishing eigenvalues of the effective connectivity matrix $\mathbf{W}$
in LIF networks, all eigenvalues $\lambda_{j}$ are represented with
their corresponding time scales in the covariances, a result we use
in \nameref{sub:Correlations-uniquely-determine-independent}.

The matrix of prefactors for the term with time dependence $\exp\left[(\lambda_{j}-1)\Delta/\tau\right]$
in expressions (55)
%\eqref{eq:binary_cov_matrix_vs_time} 
and (71)
%\eqref{eq:LIF_cov_matrix}
for the average pairwise covariances can be written as
\[
\sum_{k}\mathbf{u}^{j}\frac{\mathbf{v}^{jT}\mathbf{D}\mathbf{v}^{k}}{2-\lambda_{j}-\lambda_{k}}\mathbf{u}^{kT},
\]
where $\mathbf{D}$ is a diagonal matrix with 
\begin{eqnarray*}
\mathbf{D} & = & \begin{cases}
2\mathbf{A} & \mathrm{for}\:\mathrm{binary}\\
\frac{\lambda_{j}(2-\lambda_{j})}{\tau}\mathbf{A} & \mathrm{for}\:\mathrm{LIF}
\end{cases}.
\end{eqnarray*}
The $k$-dependence of $\lambda_{k}$ can be taken out of the sum
by reintroducing the connectivity matrix and using $\mathbf{W}^{T}\mathbf{v}^{k}=\lambda_{k}\mathbf{v}^{k}$,
\[
\sum_{k}\mathbf{u}^{j}\frac{\mathbf{v}^{jT}\mathbf{D}\mathbf{v}^{k}}{2-\lambda_{j}-\lambda_{k}}\mathbf{u}^{kT}=\mathbf{u}^{j}\mathbf{v}^{jT}\underbrace{\mathbf{D}\left[2-\lambda_{j}-\mathbf{W}^{T}\right]^{-1}}_{\equiv\left(\mathbf{B}^{-1}\right)^{T}}\underbrace{\sum_{k}\mathbf{v}^{k}\mathbf{u}^{kT}}_{\text{\ensuremath{\unit{\mathbf{1}\!\!\mathbf{1}}}}},
\]
where we have also brought the other terms that do not depend on $k$
in front of the sum, and used the biorthogonality of the left and
right eigenvectors $\mathbf{v}^{T},\mathbf{u}$ of $\mathbf{W}$.
For the time scale corresponding to $\lambda_{j}$ not to be represented,
the above expression should vanish. We show as follows that this gives
a contradiction, implying that all time scales must be represented.
Since $\mathbf{u}^{j}$ is an eigenvector, it must have at least one
nonzero entry, say for population $\alpha$. For the outer product$\left(\mathbf{u}^{j}\otimes\left[\mathbf{v}^{jT}\left(\mathbf{B}^{-1}\right)^{T}\right]\right)_{\alpha\beta}=\mathbf{u}_{\alpha}^{j}\left[\mathbf{v}^{jT}\left(\mathbf{B}^{-1}\right)^{T}\right]_{\beta}$
to vanish for all $\alpha,\beta$, the term $\left[\mathbf{v}^{jT}\left(\mathbf{B}^{-1}\right)^{T}\right]_{\beta}$
should thus vanish for all $\beta$. Both $\mathbf{B}=\mathbf{D}^{-1}\left(2-\lambda_{j}-\mathbf{W}\right)$
and $\mathbf{B}^{-1}$ are well-defined unless $\lambda_{j}=0$ in
a LIF network, or one or more populations are inactive, yielding vanishing
entries in $\mathbf{D}$. Thus, the condition for the contribution
to the covariance to vanish for all pairs of populations becomes $\mathbf{B}^{-1}\mathbf{v}^{j}=\mathbf{0}$
or $\mathbf{v}^{j}=\mathbf{B}\cdot\mathbf{0}=\mathbf{0}$, which is
inconsistent with the fact that $\mathbf{v}^{j}$ is an eigenvector.
Hence, time scales corresponding to all eigenvalues are represented
in the covariances.

\subsection{Correlations uniquely determine effective connectivity: population-independent
transfer function, $\mathbf{d=0}$\label{sub:Correlations-uniquely-determine-independent}}

In this section we show for both binary and LIF networks with population-independent
input statistics and without delays that under fairly general conditions,
the shapes of the average pairwise cross-covariances and their population
structure uniquely determine the effective connectivity. This argument
extends the one-dimensional example given in \nameref{sub:Correlations-uniquely-determine}.
As before, we assume the transfer function $H(\omega)$ itself, and
in particular the time constant $\tau$, to be unchanged under scaling.
Furthermore, we exclude the trivial scenarios where one or more of
the populations are inactive, or do not interact either with themselves
or any other population. The covariance matrix $\mathbf{c}(\Delta,d=0)$
is then given by (55)
%\eqref{eq:binary_cov_matrix_vs_time} 
for binary networks and (71)
%\eqref{eq:LIF_cov_matrix} 
for LIF networks. Since the dependence on the time interval $\Delta$ of each of these expressions
is determined by the eigenvalues $\lambda_{j}$, any scaling transformation
should keep these constant if it is to preserve the shape of the covariances.
Even for a LIF network with $\lambda_{j}=0$, where the corresponding
term drops out of the sum, this eigenvalue needs to be preserved (the
only exception being that it may become equal to another existing
eigenvalue), since otherwise an additional time dependence would
appear.  Besides $\exp[(\lambda_{j}-1)\Delta/\tau]$, the prefactor
of this term should be unchanged for each $j$ at least if there are
no degenerate or vanishing eigenvalues, as each exponential function
contributes a fall-off with a unique characteristic time scale to
the sums in (55)
%\eqref{eq:binary_cov_matrix_vs_time} 
and (71).
%\eqref{eq:LIF_cov_matrix}.
For populations $\alpha,\beta$, these prefactors can be written as
$\sum_{k}a_{jk}u_{\alpha}^{j}u_{\beta}^{kT}$ for both binary and
LIF networks, where $a_{jk}$ is a scalar that depends on $\lambda_{j}$
and $\lambda_{k}$. To preserve the population structure of the covariances
under any scaling transformation, also the ratio $\sum_{k}a_{jk}u_{\alpha_{1}}^{j}u_{\beta}^{kT}/\sum_{k}a_{jk}u_{\alpha_{2}}^{j}u_{\beta}^{kT}$
should be unchanged. As shown in \nameref{sub:Representation-of-eigenvalues},
with the exception of LIF networks with $\lambda_{j}=0$, there is
always at least one pair of populations $\alpha_{2},\beta$ with interactions
on the time scale corresponding to $\lambda_{j}$, such that this
ratio is well-defined and equals $u_{\alpha_{1}}^{j}/u_{\alpha_{2}}^{j}$.
That is, the eigenvector entries should be preserved relative to each
other, fixing the eigenvectors up to a scaling factor. Assuming that
$\mathbf{W}$ is diagonalizable, the combined conditions on the eigenvalues
and eigenvectors fix the effective connectivity matrix via $\mathbf{W}=\mathbf{U}\mathrm{diag}(\lambda_{1},\ldots,\lambda_{N_{\mathrm{pop}}})\mathbf{U}^{-1}$
where $\mathbf{U}=\left(\mathbf{u}^{1},\ldots,\mathbf{u}^{N_{\mathrm{pop}}}\right)$
is the matrix of right eigenvectors of $\mathbf{W}$.

Thus, correlation structure uniquely determines the effective connectivity
matrix at least if it is diagonalizable, and if its eigenvalues are
neither zero nor degenerate.

\paragraph*{}

\subsection{Correlations uniquely determine effective connectivity: population-independent
transfer function, general $\mathbf{d}$\label{sub:Correlations-uniquely-determine-independent-with-delays}}

Here we extend the argument of the previous sections to networks with
transmission delays. To this end, it is again convenient to work in
the Fourier domain. Since the Fourier transform is an isomorphism,
the conclusions hold also in the time domain. The rate equation (75)
%\eqref{eq:rate_dynamics_Fourier}
for the LIF dynamics can be solved for the rates $\mathbf{Y}(\omega)$
as

\begin{equation}
\mathbf{Y}(\omega)=\frac{1}{1-H(\omega)\,\mathbf{W}}\mathbf{X}(\omega),\nonumber%\label{eq:LIF_rate_eq}
\end{equation}
while for binary networks we obtain [53]
%\cite{Grytskyy13_131}
\begin{equation}
\mathbf{Y}(\omega)=\frac{H(\omega)}{1-H(\omega)\,\mathbf{W}}\mathbf{X}(\omega),\nonumber
\end{equation}
where $\mathbf{X}(\omega)$ is Gaussian white noise with amplitude
determined by the autocorrelations, and the transfer function $H(\omega)$
is given by (57).
%\eqref{eq:transfer_fn} 
For both types of networks, the
dynamics can be decomposed into eigenmodes 
\begin{equation}
\mathbf{Y}(\omega)=\sum_{k}\eta_{k}(\omega)\,\mathbf{u}^{k},\nonumber
\end{equation}
with \foreignlanguage{english}{\textrm{$\mathbf{u}^{k}$}} the right-sided
eigenvectors of $\mathbf{W}$. Let 
\begin{eqnarray*}
\gamma(\omega) & = & \begin{cases}
H(\omega) & \mathrm{for}\:\mathrm{binary}\\
1 & \mathrm{for}\:\mathrm{LIF}
\end{cases}.
\end{eqnarray*}
In terms of the left-sided eigenvectors $\mathbf{v}^{k}$ of $\mathbf{W}$,
the coefficients \foreignlanguage{english}{\textrm{$\eta_{k}(\omega)$}}
are then given by (cf. (76))
%\eqref{eq:rate_equation_projected}
\begin{eqnarray*}
\eta_{k}(\omega)=\mathbf{v}^{kT}\,\mathbf{Y}(\omega) & = & \frac{\gamma(\omega)}{1-H(\omega)\lambda_{k}}\,\mathbf{v}^{kT}\,\mathbf{X}(\omega).\nonumber
\end{eqnarray*}
Assuming that $\mathbf{W}$ has no degenerate eigenvalues and that
$\mathbf{X}(\omega)$ is nonzero for all populations (no inactive
populations), the power spectrum of each component, $\langle\eta_{k}(-\omega)\eta_{k}(\omega)\rangle\propto\left|\frac{\gamma(\omega)}{1-H(\omega)\lambda_{k}}\right|^{2}$
has a unique shape. 

As before, $\lambda_{k}=0$ in a LIF network presents a special case:
The spectrum of its coefficient reduces to a constant in the Fourier
domain, corresponding to a delta function in the time domain. This
mode only contributes to the autocovariances and not the cross-covariances,
leaving the freedom to change the corresponding right-sided eigenvector
without affecting the cross-covariances, consistent with the example
in \nameref{sub:Symmetric-two-population-spiking}.
%\textbf{"Symmetric two-population spiking network"}. 
However, such
transformations also preserve the autocovariances, despite the change
in the population structure of the contribution from the $\lambda_{k}=0$
mode. This becomes clear by rewriting the rate equation for the LIF network as

\begin{equation}
\mathbf{Y}(\omega)=\mathbf{X}(\omega)+\frac{H(\omega)\,\mathbf{W}}{1-H(\omega)\,\mathbf{W}}\mathbf{X}(\omega),\nonumber%\label{eq:LIF_rate_eq-1}
\end{equation}
showing that the part of $\bar{\mathbf{C}}(\omega)$ corresponding
to the delta peak in the time domain remains

\begin{eqnarray}
\bar{\mathbf{C}}_{\delta} & = & \langle\mathbf{X}(\omega)\mathbf{X}^{T}(-\omega)\rangle,\nonumber%\label{eq:GWN}
\end{eqnarray}
as the fact that $\mathbf{X}(\omega)$ is white noise ensures that
the expression above is just a constant, independent of $\omega$. Hence,
symmetric LIF networks where one of the eigenvalues of $\mathbf{W}$
is zero form an exception to the rule that the correlations uniquely
determine the effective connectivity.

Apart from this exception, the argument continues as follows. The
power spectra together make up the covariance matrix in the Fourier
domain
\begin{eqnarray*}
\bar{\mathbf{C}}(\omega) & = & \langle\mathbf{Y}(-\omega)\mathbf{Y}(\omega)^{T}\rangle\\
 & = & \sum_{j,k}\langle\eta_{j}(-\omega)\eta_{k}(\omega)\rangle\mathbf{u}^{j}\mathbf{u}^{kT}.
\end{eqnarray*}
If $\mathbf{W}$ is diagonalizable, the $\mathbf{u}^{k}$ are linearly
independent. Therefore, $\mathbf{u}^{k}\mathbf{u}^{kT}$ cannot be
expressed as a linear combination of the remaining terms $\mathbf{u}^{j}\mathbf{u}^{kT}$.
It thus suffices to consider the contribution of a single mode
\begin{eqnarray*}
 &  & \langle\eta_{k}(-\omega)\eta_{k}(\omega)\rangle\mathbf{u}^{k}\mathbf{u}^{kT},
\end{eqnarray*}
as its population structure makes a unique contribution to the covariance
matrix. If the covariance matrix is to be preserved, the latter term
must hence be preserved. This implies that $\lambda_{k}$ cannot change,
since it governs the covariance shape as a function of $\omega$,
and hence the temporal structure. Since $\mathbf{u}^{k}$ is by definition
an eigenvector, it has at least one non-vanishing component, say $u_{\alpha}^{k}\neq0$.
Then the $\alpha$-th row of the outer product, 
\begin{eqnarray*}
 &  & u_{\alpha}^{k}\,(u_{1}^{k},\ldots,u_{N_{\mathrm{pop}}}^{k}),
\end{eqnarray*}
must be preserved (except for $\lambda_{k}=0$ in a LIF network, as
explained above). At the $\alpha$-th column the entry is $\left(u_{\alpha}^{k}\right)^{2}$,
so $u_{\alpha}^{k}$ can only differ by a factor $\rho\in\{-1,+1\}$.
The conservation of the remaining entries $u_{\alpha}^{k}u_{\beta}^{k}$,
$\beta\neq\alpha$ implies that the $u_{\beta}^{k}$ are multiplied
by the same factor $\rho$. Hence the eigenvector must have the same
direction. As before, by the diagonalizability of $\mathbf{W}$, the
temporal and population structure of the correlations thus fix the
effective connectivity matrix via $\mathbf{W}=\mathbf{U}\mathrm{diag}(\lambda_{1},\ldots,\lambda_{N_{\mathrm{pop}}})\mathbf{U}^{-1}$
with $\mathbf{U}=\left(\mathbf{u}^{1},\ldots,\mathbf{u}^{N_{\mathrm{pop}}}\right)$
the matrix of right eigenvectors of $\mathbf{W}$.

\subsection{Correlations uniquely determine effective connectivity: population-dependent
transfer function, binary networks\label{sub:Correlations-uniquely-determine-dependent-binary}}

In \nameref{sub:Correlations-uniquely-determine-general},
%\textbf{\textcolor{blue}{"Correlations uniquely determine effective connectivity: the general case"}, 
we demonstrated
a one-to-one correspondence between the effective connectivity and
the correlations for LIF networks with non-identical populations.
We here show that the same result is obtained for binary networks
using analogous arguments. As before, we assume the summed cross-
and auto-covariance matrix in frequency domain $\bar{\mathbf{C}}(\omega)=\mathbf{C}(\omega)+\mathbf{A}(\omega)$
to be invertible, and we expand the inverse of (63)
%\eqref{eq:binary_cov_matrix_general_H}
with $\mathbf{D}(\omega)=\mathbf{H}(\omega)\mathbf{D}\mathbf{H}(-\omega)$
and $\mathbf{H}(\omega)=\mathrm{diag}\left(\left\{ H_{\alpha}(\omega)\right\} _{\alpha=1\ldots N_{\mathrm{pop}}}\right)$
to obtain the diagonal element

\begin{eqnarray}
\bar{C}_{\alpha\alpha}^{-1} & = & \frac{1+\omega^{2}\tau_{\alpha}^{2}}{D_{\alpha}}\nonumber \\
 & - & \frac{W_{\alpha\alpha}}{D_{\alpha}}\left(e^{-i\omega d_{\alpha\alpha}}(1-i\omega\tau_{\alpha})+e^{i\omega d_{\alpha\alpha}}(1+i\omega\tau_{\alpha})\right)\nonumber \\
 & + & \sum_{\gamma}\frac{W_{\gamma\alpha}^{2}}{D_{\gamma}}.\nonumber
\end{eqnarray}
Since $D_{\alpha}^{-1}$ determines a quadratic dependence on $\omega$
that cannot be offset by other terms, it needs to be preserved. This
fixes $W_{\alpha\alpha}$, which similarly determines a unique $\omega$-dependence.
Furthermore, we have for $\alpha\neq\beta$

\begin{eqnarray}
\bar{C}_{\alpha\beta}^{-1} & = & \frac{W_{\alpha\beta}}{D_{\alpha}}e^{-i\omega d_{\alpha\beta}}\left(-1+i\omega\tau_{\alpha}+W_{\alpha\alpha}e^{i\omega d_{\alpha\alpha}}\right)\nonumber \\
 & + & \frac{W_{\beta\alpha}}{D_{\beta}}e^{i\omega d_{\beta\alpha}}\left(-1-i\omega\tau_{\beta}+W_{\beta\beta}e^{-i\omega d_{\beta\beta}}\right)\nonumber \\
 & + & \sum_{\gamma\neq\alpha,\beta}\frac{W_{\gamma\alpha}W_{\gamma\beta}}{D_{\gamma}}e^{i\omega\left(d_{\gamma\alpha}-d_{\gamma\beta}\right)}.\nonumber
\end{eqnarray}
Here, the term $W_{\alpha\beta}D_{\alpha}^{-1}e^{-i\omega d_{\alpha\beta}}i\omega\tau_{\alpha}$
cannot be offset by other terms unless $d_{\alpha\beta}=d_{\beta\alpha}=0$,
showing that $W_{\alpha\beta}$ needs to be unchanged in order to
keep $\bar{C}_{\alpha\beta}^{-1}$ constant. In contrast to the LIF
case, $\mathbf{C}(\omega)$ differs from $\bar{\mathbf{C}}(\omega)$
not by constant terms, but by $\mathrm{diag}\left(\left\{ \frac{2\tau_{\alpha}}{1+\omega^{2}\tau_{\alpha}^{2}}\frac{a_{\alpha}}{N_{\alpha}}\right\} _{\alpha=1\ldots N_{\mathrm{pop}}}\right)$.
Therefore, a priori it appears that there may be a freedom to scale
both the population sizes $N_{\alpha}$ and terms in $\bar{\mathbf{C}}(\omega)$
with the same inverse quadratic $\omega$-dependence. We can see what
this entails by considering
\begin{eqnarray}
\bar{\mathbf{Q}}(\omega) & \equiv & \bar{\mathbf{C}}(\omega)\mathrm{diag}\left(\left\{ 1+\omega^{2}\tau_{\alpha}^{2}\right\} _{\alpha=1\dots N_{\mathrm{pop}}}\right)\nonumber \\
 & = & \mathbf{C}(\omega)\mathrm{diag}\left(\left\{ 1+\omega^{2}\tau_{\alpha}^{2}\right\} _{\alpha=1\dots N_{\mathrm{pop}}}\right)\nonumber \\
 & + & \mathrm{diag}\left(\left\{ \frac{2\tau_{\alpha}a_{\alpha}}{N_{\alpha}}\right\} _{\alpha=1\dots N_{\mathrm{pop}}}\right).\nonumber
\end{eqnarray}
This shows that changing any $\omega$-dependent terms in $\bar{\mathbf{Q}}(\omega)$
would change the $\omega$-dependence of $\mathbf{C}(\omega)$. Furthermore,
the elements of $\bar{\mathbf{Q}}^{-1}(\omega)$ have the same form
as those of $\bar{\mathbf{C}}^{-1}(\omega)$ for the LIF network except
for the index of $\tau_{\alpha}$, with diagonal elements
\begin{align}
\bar{Q}_{\alpha\alpha}^{-1}(\omega) & =\frac{1}{D_{\alpha}}\nonumber \\
 & -\frac{W_{\alpha\alpha}}{D_{\alpha}}\left(\frac{e^{-i\omega d_{\alpha\alpha}}}{1+i\omega\tau_{\alpha}}+\frac{e^{i\omega d_{\alpha\alpha}}}{1-i\omega\tau_{\alpha}}\right)\nonumber \\
 & +\sum_{\gamma}\frac{1}{D_{\gamma}}\frac{W_{\gamma\alpha}^{2}}{1+\omega^{2}\tau_{\alpha}^{2}},\nonumber
\end{align}
and off-diagonal elements
\begin{eqnarray}
\bar{Q}_{\alpha\beta}^{-1} & = & \frac{W_{\alpha\beta}}{D_{\alpha}}e^{-i\omega d_{\alpha\beta}}\left(-\frac{1}{1+i\omega\tau_{\alpha}}+W_{\alpha\alpha}\frac{e^{i\omega d_{\alpha\alpha}}}{1+\omega^{2}\tau_{\alpha}^{2}}\right)\nonumber \\
 & + & \frac{W_{\beta\alpha}}{D_{\beta}}e^{i\omega d_{\beta\alpha}}\left(-\frac{1}{1+\omega^{2}\tau_{\alpha}^{2}}(1+i\omega\tau_{\beta})+W_{\beta\beta}\frac{e^{-i\omega d_{\beta\beta}}}{1+\omega^{2}\tau_{\alpha}^{2}}\right)\nonumber \\
 & + & \sum_{\gamma\neq\alpha,\beta}\frac{W_{\gamma\alpha}W_{\gamma\beta}}{D_{\gamma}}\frac{e^{i\omega\left(d_{\gamma\alpha}-d_{\gamma\beta}\right)}}{1+\omega^{2}\tau_{\alpha}^{2}}.\nonumber
\end{eqnarray}
Hence, comparing to (14),
%\eqref{eq:C_inv_LIF_alpha_beta}
we reach the same conclusion as for the LIF network: in order to preserve $\mathbf{C}(\omega)$,
$\mathbf{D}$ and $\mathbf{W}$ must not change, at least if all connections
exist, and if there are no symmetries in the delays and time constants
like those described in \nameref{sub:Correlations-uniquely-determine-general}.
%\textbf{\textcolor{blue}{"Correlations uniquely determine effective connectivity: the general case"}.

\end{document}